\documentclass[preprint2,longabstract]{aastex}

\slugcomment{}

\shorttitle{The all-particle spectrum of primary cosmic rays}
\shortauthors{M. Amenomori et al.}

\begin{document}

%% LaTeX will automatically break titles if they run longer than
%% one line. However, you may use \\ to force a line break if
%% you desire.

\title{The all-particle spectrum of primary cosmic rays \\
in the wide energy range from $10^{14}$ eV to $10^{17}$ eV \\ 
observed with the Tibet-III air-shower array
}

%% Use \author, \affil, and the \and command to format
%% author and affiliation information.
%% Note that \email has replaced the old \authoremail command
%% from AASTeX v4.0. You can use \email to mark an email address
%% anywhere in the paper, not just in the front matter.
%% As in the title, use \\ to force line breaks.

\author{
M.~Amenomori\altaffilmark{1},
X.~J.~Bi\altaffilmark{2},
D.~Chen\altaffilmark{3},
S.~W.~Cui\altaffilmark{4},
Danzengluobu\altaffilmark{5},
L.~K.~Ding\altaffilmark{2},
X.~H.~Ding\altaffilmark{5},
C.~Fan\altaffilmark{6},
C.~F.~Feng\altaffilmark{6},
Zhaoyang Feng\altaffilmark{2},
Z.~Y.~Feng\altaffilmark{7},
X.~Y.~Gao\altaffilmark{8},
Q.~X.~Geng\altaffilmark{8},
H.~W.~Guo\altaffilmark{5},
H.~H.~He\altaffilmark{2},
M.~He\altaffilmark{6},
K.~Hibino\altaffilmark{9},
N.~Hotta\altaffilmark{10},
Haibing~Hu\altaffilmark{5},
H.~B.~Hu\altaffilmark{2},
J.~Huang\altaffilmark{11},
Q.~Huang\altaffilmark{7},
H.~Y.~Jia\altaffilmark{7},
F.~Kajino\altaffilmark{12},
K.~Kasahara\altaffilmark{13},
Y.~Katayose\altaffilmark{3},
C.~Kato\altaffilmark{14},
K.~Kawata\altaffilmark{11},
Labaciren\altaffilmark{5},
G.M.~Le\altaffilmark{15},
A.~F.~Li\altaffilmark{6},
J.Y.~Li\altaffilmark{6},
Y.-Q.~Lou\altaffilmark{16},
H.~Lu\altaffilmark{2},
S.~L.~Lu\altaffilmark{2},
X.~R.~Meng\altaffilmark{5},
K.~Mizutani\altaffilmark{13,17},
J.~Mu\altaffilmark{8},
K.~Munakata\altaffilmark{14},
A.~Nagai\altaffilmark{18},
H.~Nanjo\altaffilmark{1},
M.~Nishizawa\altaffilmark{19},
M.~Ohnishi\altaffilmark{11},
I.~Ohta\altaffilmark{20},
H.~Onuma\altaffilmark{17},
T.~Ouchi\altaffilmark{9},
S.~Ozawa\altaffilmark{11},
J.~R.~Ren\altaffilmark{2},
T.~Saito\altaffilmark{21},
T.~Y.~Saito\altaffilmark{22},
M.~Sakata\altaffilmark{12},
T.~K.~Sako\altaffilmark{11},
M.~Shibata\altaffilmark{3},
A.~Shiomi\altaffilmark{9,11},
T.~Shirai\altaffilmark{9},
H.~Sugimoto\altaffilmark{23},
M.~Takita\altaffilmark{11},
Y.~H.~Tan\altaffilmark{2},
N.~Tateyama\altaffilmark{9},
S.~Torii\altaffilmark{13},
H.~Tsuchiya\altaffilmark{24},
S.~Udo\altaffilmark{11},
B.~Wang\altaffilmark{8},
H.~Wang\altaffilmark{2},
X.~Wang\altaffilmark{11},
Y.~Wang\altaffilmark{2},
Y.~G.~Wang\altaffilmark{6},
H.~R.~Wu\altaffilmark{2},
L.~Xue\altaffilmark{6},
Y.~Yamamoto\altaffilmark{12},
C.~T.~Yan\altaffilmark{11},
X.~C.~Yang\altaffilmark{8},
S.~Yasue\altaffilmark{25},
Z.~H.~Ye\altaffilmark{15},
G.~C.~Yu\altaffilmark{7},
A.~F.~Yuan\altaffilmark{5},
T.~Yuda\altaffilmark{9},
H.~M.~Zhang\altaffilmark{2},
J.~L.~Zhang\altaffilmark{2},
N.~J.~Zhang\altaffilmark{6},
X.~Y.~Zhang\altaffilmark{6},
Y.~Zhang\altaffilmark{2},
Yi.~Zhang\altaffilmark{2},
Zhaxisangzhu\altaffilmark{5},
and X.~X.~Zhou\altaffilmark{7}, \\
(The Tibet AS$\gamma$ Collaboration)}

\altaffiltext{1}{Department of Physics, Hirosaki University, Hirosaki 036-8561, Japan.}
\altaffiltext{2}{Key Laboratory of Particle Astrophysics, Institute of High Energy Physics, Chinese Academy of Sciences, Beijing 100049, China. }
\altaffiltext{3}{Faculty of Engineering, Yokohama National University, Yokohama 240-8501, Japan. }
\altaffiltext{4}{Department of Physics, Hebei Normal University, Shijiazhuang 050016, China. }
\altaffiltext{5}{Department of Mathematics and Physics, Tibet University, Lhasa 850000, China. }
\altaffiltext{6}{Department of Physics, Shandong University, Jinan 250100, China. }
\altaffiltext{7}{Institute of Modern Physics, SouthWest Jiaotong University, Chengdu 610031, China. }
\altaffiltext{8}{Department of Physics, Yunnan University, Kunming 650091, China.}
\altaffiltext{9}{Faculty of Engineering, Kanagawa University, Yokohama 221-8686, Japan.}
\altaffiltext{10}{Faculty of Education, Utsunomiya University, Utsunomiya 321-8505, Japan. }
\altaffiltext{11}{Institute for Cosmic Ray Research, University of Tokyo, Kashiwa 277-8582, Japan.}
\altaffiltext{12}{Department of Physics, Konan University, Kobe 658-8501, Japan.}
\altaffiltext{13}{Research Institute for Science and Engineering, Waseda University, Tokyo 169-8555, Japan.}
\altaffiltext{14}{Department of Physics, Shinshu University, Matsumoto 390-8621, Japan.}
\altaffiltext{15}{Center of Space Science and Application Research, Chinese Academy of Sciences, Beijing 100080, China.}
\altaffiltext{16}{Physics Department and Tsinghua Center for Astrophysics, Tsinghua University, Beijing 100084, China.}
\altaffiltext{17}{Department of Physics, Saitama University, Saitama 338-8570, Japan.}
\altaffiltext{18}{Advanced Media Network Center, Utsunomiya University, Utsunomiya 321-8585, Japan.}
\altaffiltext{19}{National Institute of Informatics, Tokyo 101-8430, Japan.}
\altaffiltext{20}{Tochigi Study Center, University of the Air, Utsunomiya 321-0943, Japan.}
\altaffiltext{21}{Tokyo Metropolitan College of Industrial Technology, Tokyo 116-8523, Japan.}
\altaffiltext{22}{Max-Planck-Institut f\"ur Physik, M\"unchen D-80805, Deutschland.}
\altaffiltext{23}{Shonan Institute of Technology, Fujisawa 251-8511, Japan.}
\altaffiltext{24}{RIKEN, Wako 351-0198, Japan.}
\altaffiltext{25}{School of General Education, Shinshu University, Matsumoto 390-8621, Japan.}

%% Notice that each of these authors has alternate affiliations, which
%% are identified by the \altaffilmark after each name.  Specify alternate
%% affiliation information with \altaffiltext, with one command per each
%% affiliation.

%% Mark off your abstract in the ``abstract'' environment. In the manuscript
%% style, abstract will output a Received/Accepted line after the
%% title and affiliation information. No date will appear since the author
%% does not have this information. The dates will be filled in by the
%% editorial office after submission.

\begin{abstract}
We present an updated all-particle energy spectrum of primary cosmic rays
in a wide range from $10^{14}$ eV to $10^{17}$ eV using $5.5 \times 10^{7}$ 
events collected in the period from 2000 November through 2004 October by 
the Tibet-III air-shower array located at 4300 m above sea level
(atmospheric depth of 606 g/cm$^{2}$). The size spectrum exhibits a sharp knee 
at a corresponding primary energy around 4 PeV. This work uses increased 
statistics and new simulation calculations for the analysis.
We performed extensive Monte Carlo calculations and  discuss the model 
dependences involved in the final result assuming interaction models of 
QGSJET01c and SIBYLL2.1 and primary composition models 
of heavy dominant (HD) and proton dominant (PD) ones.
Pure proton and pure iron primary models are also examined as extreme
cases. The detector simulation was also made to improve the accuracy of 
determining the size of the air showers and the energy of the primary
particle. We confirmed that the all-particle energy spectra obtained
under various plausible model parameters are not significantly 
different from each other as expected from the characteristics of 
the experiment at the high altitude, where the air showers of the primary 
energy around the knee reaches near maximum development and their features
are dominated by electromagnetic components leading to the weak dependence 
on the interaction model or the primary mass. This is the highest-statistical 
and the best systematics-controlled measurement covering the widest energy 
range around the knee energy region.

\end{abstract}

%% Keywords should appear after the \end{abstract} command. The uncommented
%% example has been keyed in ApJ style. See the instructions to authors
%% for the journal to which you are submitting your paper to determine
%% what keyword punctuation is appropriate.

\keywords{cosmic rays---methods: data analysis---stars: supernovae : general}

\section{Introduction}
\label{intro}

Although almost 100 years have passed over since the discovery of cosmic rays, their source and 
acceleration mechanism are still not fully understood. 
The energy spectrum and chemical composition of cosmic rays can be key
information for probing their origin, acceleration mechanism and propagation 
mechanism. The cosmic-ray spectrum has been observed by many ground-based experiments 
to resemble two power laws,
having a form dj/dE $\propto$ $E^{-\gamma}$, with $\gamma$ = 2.7 below
the energy around 4 $\times$ $10^{15}$ eV, and then steepening to $\gamma$ = 3.1 above 
this energy \cite{Hoerandel2}. The change of the power index at this energy
is called the spectral ``knee''.  Although the existence of the knee has been well 
established experimentally, there are still controversial arguments on its origin. 
Proposals for its origin range from astrophysical scenarios like the change of 
acceleration mechanisms \citep{Berezhko,Stanev,Kobayakawa,Volk}
at the sources of cosmic rays (supernova remnants, 
pulsars, etc.), the single source assumption \cite{Erlykin}, 
or effects due to the propagation \citep{Ptuskin,Candia} 
inside the galaxy (diffusion, drift, escape from the Galaxy) to 
particle-physics models like the interaction with relic 
neutrinos \cite{Wigmans} during transport or new processes in the 
atmosphere \cite{Nikolsky} during air-shower development. 
Common to all models is the prediction of a change of the chemical composition 
over the knee region. 
Direct measurements of primary cosmic rays on board balloons or satellites 
are the best ways for the study of the chemical composition, 
however, the energy region covered by them with sufficient statistics
are limited to $10^{14}$ eV. The energy spectrum and chemical composition of 
primary cosmic rays around the knee, therefore, has to be studied with 
ground-based air-shower experiments using surface array and/or 
detectors for Cerenkov light.

Many reports have so far been made on the energy spectrum as well as 
the chemical composition of primary cosmic rays. 
Although the global features of the all-particle spectrum agree well
when we take into account the systematic
errors of about 20\% involved in the energy scale
\cite{Hoerandel2}, there are still serious disagreements in the chemical 
composition depending on experimental methods, for example, the knee
composition obtained by the Tibet and KASCADE experiments can be summarized
as follows.
We have already reported the energy spectrum of protons and helium in
the energy range from 200 TeV to 10,000 TeV 
\citep{Amenomori5,Amenomori2} from air-shower core observation,
suggesting a steep power index of approximately -3.1. This indicates that the power
index of light component is changed from approximately -2.7 as measured 
by direct observations to -3.1 around a few hundred TeV.
Hence, the light component should become
less abundant at the knee and the main component responsible to the structure
of the knee must be heavier than helium. Furthermore, the spectral shape of
light component seems to keep the power law 
instead of the exponential cutoff. 
On the contrary, the KASCADE using electron-muon size analysis
\cite{Antoni} claims that the knee in the all-particle 
spectrum is due to the steepening of the spectra of light elements
with exponential type cutoff.

The accurate measurement of the all-particle energy spectrum around the knee
is essential to establish the chemical composition at this
energy range.  There is no precise measurement of the chemical composition
around the knee region yet and it is in fact impossible to discriminate 
individual elements clearly by indirect observations. Therefore, 
most of the works made so far simply discussed
the average mass $<\ln A>$. Another approach is the unfolding of the
all-particle spectrum using shower characteristics like electron-muon
ratio, depth of the shower maximum and so on. 
In these methods, the detailed information of the all-particle spectrum 
plays an important role in determining the chemical composition.
It is also expected that
the specific features of each component like cutoff energy or 
source characteristics should be reflected in the shape of the
all-particle spectrum as discussed in the single source model \cite{Erlykin}.
The important features of the all-particle spectrum are the absolute intensity,
the position of the knee, the difference of the power index before and
after the knee and the sharpness in the size spectrum, which are deeply
connected with the acceleration mechanism and the source of cosmic rays.

\begin{figure}
\epsscale{.80}
\plotone{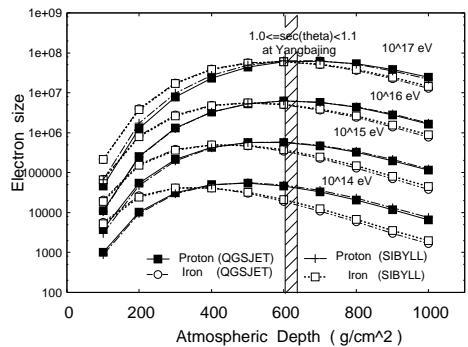}
\caption{Average transition curves of air-shower size induced by protons and iron nuclei
for a vertical incidence.
}
\label{fig:1}
\end{figure}

The merit of the air-shower experiment in Tibet is that the atmospheric 
depth of the experimental site (4300 m a.s.l., 606 g/{cm}$^{2}$) is close 
to the maximum development of the air showers with energies around the knee 
almost independent of the masses of primary cosmic rays as demonstrated in
Fig.~\ref{fig:1} for vertically incident cosmic rays. 
It should be also noted that the number of shower
particles is dominated by electromagnetic component with
minor contribution of muons, whose interaction model dependence is
known to be rather large among current interaction models leading to 
a large systematic error in the experiments carried out at the sea level 
because of the large contribution of muons.
In other words, the air shower observation at high altitude is sensitive
to the most forward region of the hadronic interactions in the center of
momentum system (CMS) where high energy secondaries are produced, and the 
electromagnetic component as a decay product of neutral pions
dominate the number of shower particles, while it is insensitive
to the central region of the CMS where large number of muons as decay product of
charged pions are produced. The differences among current
interaction models are mainly related to the central region as seen in
the problem of the electron-muon correlation.
Hence, the air-shower experiment in Tibet
can determine the primary cosmic-ray energy much less dependently 
upon the chemical composition and the interaction model
than experiments at the sea level.

%
% Fig.1
%

We have already reported the first result on the all-particle spectrum 
around the knee region based on data from 2000 November to 2001 October 
observed by the Tibet-III air-shower array \cite{Amenomori8}. 
In this paper,  
we present an updated all-particle energy spectrum using data set collected 
in the period from 2000 November through 2004 October.
The updates are due to 
(1) increased statistics by approximately 2.6 times, 
(2) use of new simulation codes and 
(3) improvement of the lateral structure function
used for the size estimation of air showers.
The previous result was obtained using almost the same  
analysis as used in Tibet I \cite{Amenomori1} except for the parameters
which depend on the detector configuration.
In the present paper, the simulation code Cosmos is replaced by
Corsika with interaction
models QGSJET01c and SIBYLL2.1, which is now widely used in many
analyses by other works and enables the comparison of this work with
others easier.
The third update on the structure function is made
to cover wider energy range than before (see section \ref{ModNKG}).

Thus, we obtained the all-particle  energy spectrum of cosmic rays 
in a wide range over 3 decades between $10^{14}$ eV  and $10^{17}$ eV 
and the updated result is compared with previous ones.

\section{Tibet experiment}

%
% Fig.2
%
\begin{figure}
\epsscale{.80}
\plotone{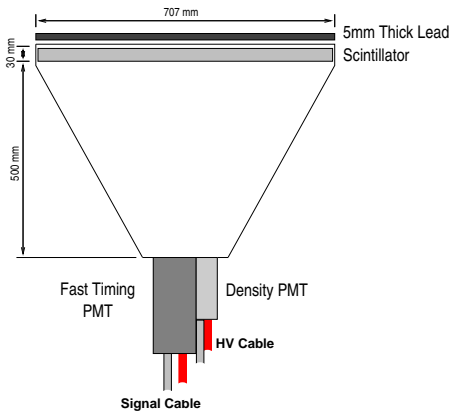}
\caption{Schematic view of the FT w/D-detector.
}
\label{fig:2}
\end{figure}

%
% Fig.3
%
\begin{figure}
\epsscale{.80}
\plotone{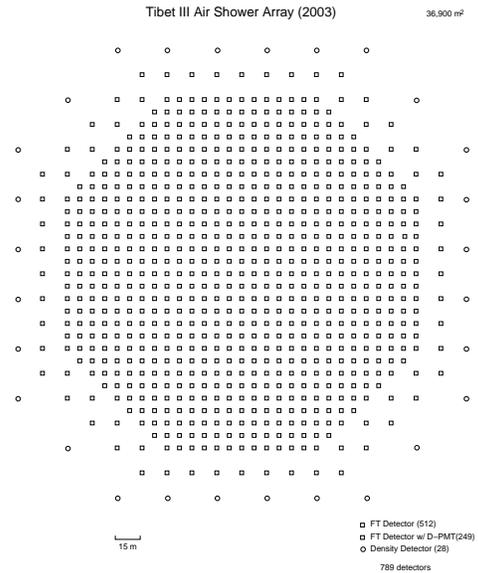}
\caption{Schematic view of the Tibet-III array operating at Yangbajing. 
The Tibet-III array consists of 761 FT detectors 
and 28 D detectors around them. In the inner 36,900 m$^2$, FT detectors 
are deployed at 7.5 m lattice intervals, 
among 761 FT counters, 249 sets of detectors are also equipped with
D-PMT in addition to FT-PMTs.
Open-white squares: FT detectors with FT-PMT; 
Open-black squares: FT detectors with FT-PMT and D-PMT; 
Open circles: density detectors only with D-PMT. 
}
\label{fig:3}
\end{figure}

The Tibet air-shower experiment has been operated
at Yangbajing (E$90^{\circ}$31', N$30^{\circ}$06';
4300 m above sea level) in Tibet, China, since 1990. 
The Tibet air-shower array is designed not only for observation of air showers of
nuclear-component origin but also for that of high energy celestial gamma rays. 
Because of such multiple purposes, the detector is constructed to cover 
a wide dynamic range for particle density covering 0.1 to 5000 and
a good angular resolution for the arrival direction of air showers with energy
in excess of a few TeV being better than 1 degree.

The Tibet-I surface array was constructed in 1990 \cite{Amenomori3}
using 65 plastic scintillation detectors placed on a lattice with 15 m spacing.
This array was gradually expanded to the Tibet-II (1994) and
Tibet-III (1999) array. At present, it consists of 761 fast timing (FT) counters 
and 28 density (D) counters surrounding them. In the inner 36,900 m$^2$, 
FT counters are deployed at 7.5 m lattice intervals. 
All the FT counters are equipped with a fast-timing photomultiplier 
tube (FT-PMT ; Hamamatsu H1161) measuring up to 15 particles.
Among the 761 FT counters, 249 sets of detectors (with interval of 15 m)
are also equipped with density photomultiplier tube (D-PMT ; Hamamatsu H3178) 
of wide dynamic range measuring up to 5000 particles in addition to FT-PMTs, 
so that UHE cosmic rays with energy above the knee can be observed 
with a good accuracy.

Each counter has a plastic scintillator
plate (BICRON BC-408A) of 0.5 m$^2$ in area and 3 cm in thickness. 
A lead plate of 0.5 cm thick is put on the top of each counter 
as shown in Fig.~\ref{fig:2} in order to increase 
the counter's sensitivity by converting photons
in an electromagnetic shower into electron-positron pairs 
\citep{Bloomer,Amenomori4}.
The recording of signals is made on
time and charge information for the FT-PMTs,
while only the charge information for the D-PMTs.
The D counters surrounding the inner array
are also equipped with both FT-PMT and D-PMT, where only the charge information 
of both PMTs are recorded. 
An event trigger signal is issued when any fourfold 
coincidence occurs in the FT counters recording more than 0.6 particles.
Fig.~\ref{fig:3} is the schematic view of the Tibet-III array.

The primary energy of each event is determined by the shower 
size $N_e$, which is calculated by fitting the lateral particle density 
distribution to the modified NKG structure function (see section \ref{ModNKG}). 
The air-shower direction can be estimated 
with an inaccuracy smaller than 0.2$^{\circ}$ at energies above 10$^{14}$ eV,
which is calibrated by observing the Moon's shadow \cite{Amenomori9}.
We used the data set obtained during the period from 2000 November through 
2004 October. The effective live time used for the present analysis is 805.17 days.

\section{Simulation}
 Monte Carlo simulation (MC) plays an important role in air-shower experiments 
since most of the methods of the analysis are developed so that they can reproduce 
the inputs of simulated events like the primary energy, the location of the shower axis, 
the arrival direction and so on. Even the most basic quantity like the number of 
particles arriving to a detector should be 'defined' through MC because we do not 
measure the number of particles but the charge of PMT-output which is not simply
proportional to the number of charged particles entering into the detector if we take
into account the contribution of the electromagnetic processes by photons 
inside the detectors. Another example of the role of MC is to define the effective 
area of the shower array which should be determined to avoid the erroneous counting
of the events whose shower axises are dropping outside of the effective area.
Therefore detailed MC calculations are needed on air-shower generation in the
atmosphere and on the detector response. Consequently, the final result inevitably 
depends on the interaction model and on the primary composition model in MC. 
This is the main source of the systematic errors involved in the air-shower 
experiment and we try to show them explicitly in the present work.

A full Monte Carlo (MC) simulation has been carried out on the development 
of air showers in the atmosphere and also on the detector response of the 
Tibet-III array. The simulation code  CORSIKA (version 6.204) including 
QGSJET01c and SIBYLL2.1 interaction models \cite{Heck1}
is used to generate air-shower events. All shower particles in the atmosphere
are traced down to the minimum energy of 1 MeV without using thinning method. 

\begin{figure}
\begin{minipage}{8.cm}
\begin{center}
\plotone{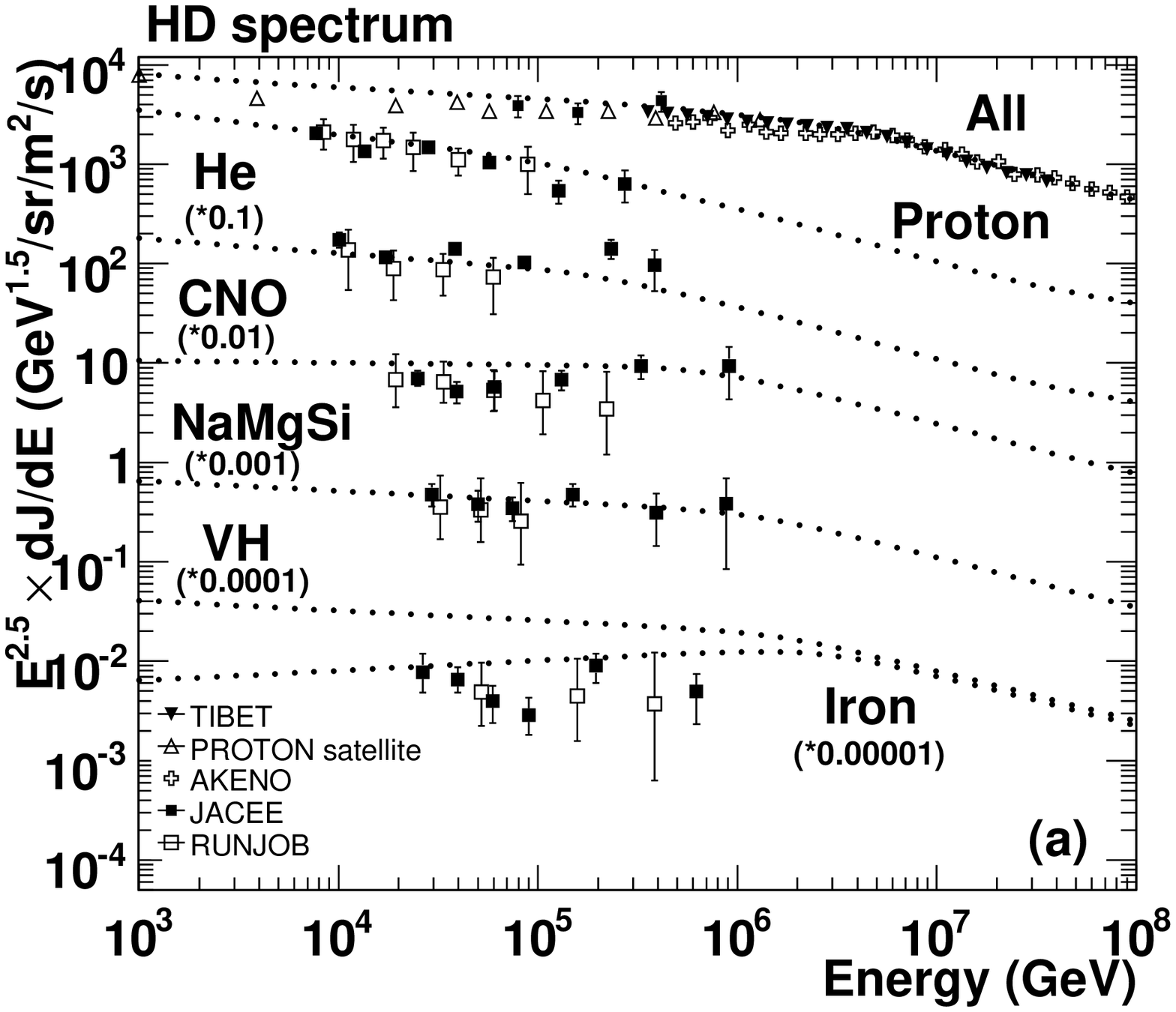}
\end{center}
\end{minipage}
\begin{minipage}{8.cm}
\begin{center}
\plotone{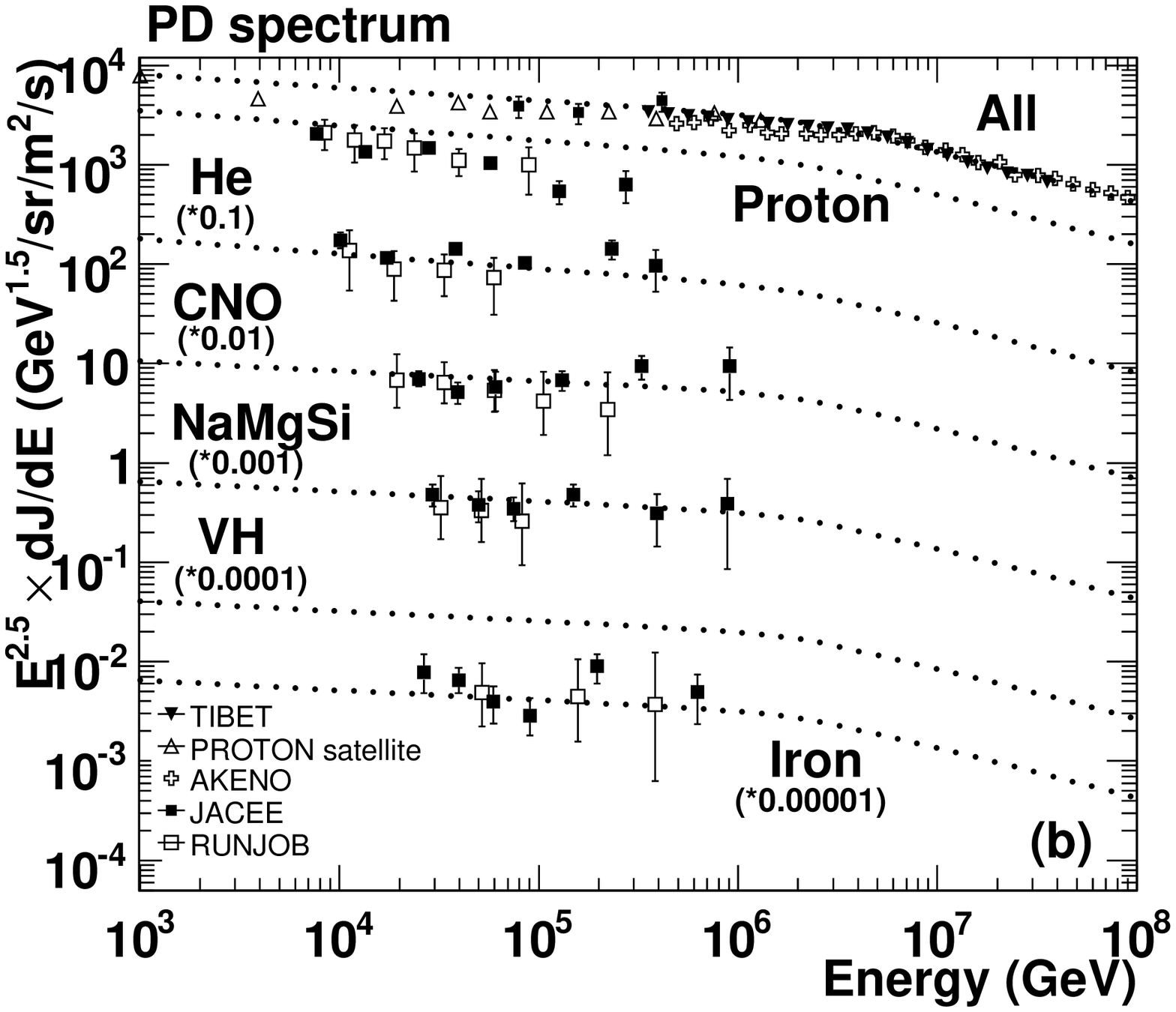}
\end{center}
\end{minipage}
\caption{Primary cosmic ray composition for (a) the HD model and (b) the PD model.
The all-particle spectrum, which is a sum of each component, is normalized to
the Tibet data, and they are compared with other experiments.
PROTON satellite \cite{Grigorov}, 
AKENO \cite{Nagano1},JACEE \cite{Asakimori},RUNJOB \cite{Apanasenko}.
}
\label{fig:4}
\end{figure}

Although the chemical composition of the primary particles around the
knee region is not well established yet, we have to assume it in the
simulation.
The simplest way to bracket all possibilities is to assume pure
proton and pure iron primaries. Since it is almost evident that
such assumptions are not realistic and lead to unacceptable results 
showing disagreement with the direct observations, these results
will be mentioned as extreme cases.
More realistic treatment of the chemical composition
is to extrapolate the known composition
at low energies measured by direct observations.
The uncertainty in extrapolating to the high energy range can be treated 
by bracketing the reported results on the composition
study around the knee. 
In order to examine the composition dependence 
involved in the all-particle spectrum,
we used two kinds of the mixed composition models.
One is based on the dominance of heavy component around the knee 
as reported by works \citep{Amenomori2,Amenomori5,Ogio} and 
this composition is called HD model.
Another is based on the dominance of
light component (p and He) as reported by works \citep{Antoni,EASTOP,CASA-BLANCA}
and called PD model.
The energy spectra of individual mass groups in HD model and PD model
are shown in Fig.~\ref{fig:4}(a) and Fig.~\ref{fig:4}(b), respectively.
Table~\ref{tab:00} shows their fractional contents at given energies.
In total, four kind of the primary composition, namely, pure proton, pure iron,
HD model and PD model are used in the simulation with the minimum primary
energy of 50 TeV.

%
% Fig.4
%

%
% TABLE 0.
%
\begin{table}[t]
\begin{center}
\caption{Fractions of the proton(P), helium(He), CNO(M), NaMgSi(H),SClAr(VH) and iron(Fe) 
components in the assumed primary cosmic-ray spectrum of the HD and PD models \cite{Amenomori5}.}
\begin{tabular}{l|c|c|c}
\tableline\tableline
HD model & $10^{14}$-$10^{15}$ eV & $10^{15}$-$10^{16}$ eV & $10^{16}$-$10^{17}$ eV \\
\tableline
P & 22.6\%  & 11.0\% & 8.1\% \\
He & 19.2\%  & 11.4\% & 8.4\% \\
M(CNO) & 21.0\%  & 22.6\% & 17.8\% \\
H(NaMgSi) & 9.0\%  & 9.4\% & 8.1\% \\
VH(SClAr) & 5.6\%  & 6.2\% & 5.8\% \\
Fe & 22.2\% & 39.1\% & 51.7\% \\
\tableline
\tableline
PD model & $10^{14}$-$10^{15}$ eV & $10^{15}$-$10^{16}$ eV & $10^{16}$-$10^{17}$ eV \\
\tableline
P & 39.0\%  & 38.1\% & 37.5\% \\
He & 20.4\%  & 19.4\% & 19.1\% \\
M(CNO) & 15.2\%  & 16.1\% & 16.5\% \\
H(NaMgSi) & 9.4\%  & 9.9\% & 10.2\% \\
VH(SClAr) & 5.8\%  & 6.2\% & 6.3\% \\
Fe & 9.4\% & 9.9\% & 10.2\% \\
\tableline\tableline
\end{tabular}
\label{tab:00}
\end{center}
\end{table}

%
% Fig.5
%
\begin{figure}
\epsscale{.70}
\plotone{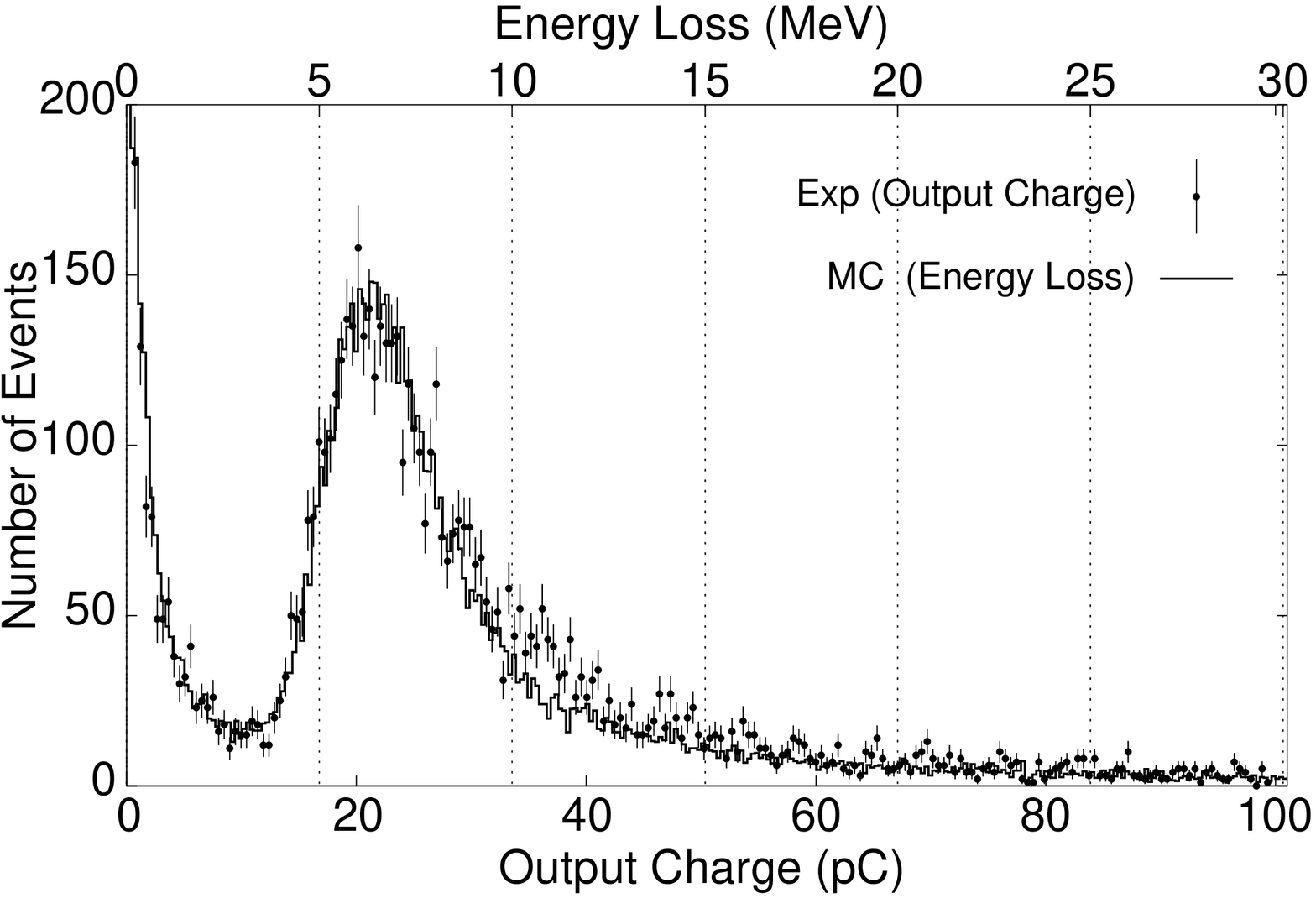}
\caption{Charge distribution in a detector measured by probe calibration
(see the text).
In order to compare the charge distribution with the simulation on 
the energy loss in a scintillator, 
the MC result is adjusted by multiplying a constant to meet with 
the same peak position as the experiment. The fluctuation of the number 
of photons in scintillation light is taken into account with the normal 
distribution in MC.
}
\label{fig:5}
\end{figure}

All secondary particles are traced until their energies become 1 MeV in 
the atmosphere. 
The shower axis was placed on Tibet array at random within a
radius of 100 m from the center of the array.
In order to treat the MC events in the same way as experimental data analysis,
simulated air-shower events were input to the detector with 
the same detector configuration as the Tibet-III array with use of Epics 
code (ver. 8.64) \cite{Kasahara} to calculate the energy deposit of these 
shower particles.
Experimentally, the number of charged particles is defined as
the PMT output (charge) divided by that of the single particle peak,
which is determined by a probe calibration using cosmic rays, 
typically single muons. 
For this purpose, a small scintillator of 25 cm $\times$ 25 cm $\times$ 3.5 cm thick 
with a PMT (H1949) is put on the top of the each detector during the maintenance period. 
This is called a probe detector and is used for making the trigger of the each Tibet-III detector. 
The response of each detector is calibrated every year through probe calibration.
In the simulation, these events triggered by the probe detector was 
also examined by a MC calculation, in which
the primary particles were sampled in the energy range above the geomagnetic 
cutoff energy at Yangbajing ($>$10 GeV), and all secondary particles 
which pass  the probe detector and the Tibet-III detector simultaneously were 
selected for the analysis. Since the value of PMT output is proportional 
to the energy loss of the particles passing through the scintillator, 
the peak position of the energy loss distribution 
corresponds  to the experimental single peak of the probe calibration.

According to the MC, the peak position of the energy loss in the scintillator
is calculated as 6.11 MeV (the details are written 
in the paper \cite{Amenomori6}). We can then calculate the number of 
charged particles for each detector hit as the total energy loss in each scintillator
divided by 6.11 MeV instead of counting the number of charged particles arriving 
to the detector in MC events. 
We confirmed that the shape of the energy loss distribution, which is determined
by probe-calibration simulation, shows a reasonable agreement with 
the charge distribution of the experimental data as shown 
in Fig.~\ref{fig:5}, where the proportionality between the energy 
loss $\Delta E$ and the PMT output charge $Q_i$ is assumed as 
$Q_i = k_i \times \Delta E$, 
where $k_i$ is a proportional constant depending on given detector
typically being around 4 pC/MeV. Thus, all detector responses including muons and 
the materialization of photons inside the detector are taken into account. 
The total number of charged particles of each event (hereafter, 
we call this an ^^ ^^  air shower size (Ne)'') 
was estimated using the modified NKG lateral distribution function 
which is tuned to reproduce the above defined number of particles
using the Monte Carlo simulation under our detector configurations. 
The number of MC events, typically for QGSJET+HD, is as follows.
About 10 million of air showers are generated with primary energy above
50 TeV. After imposing the selection criteria described
in sec.~\ref{data-selection}, 
the surviving  number of events is about 5 million, among which
1.9 million events belong to the unbiased energy region corresponding 
to E0$>$100 TeV. Almost the same number of MC events are obtained
for other models to compare with each other.

%
% Fig.6(a),(b)
%
\begin{figure}
\begin{minipage}{8.cm}
\begin{center}
\plotone{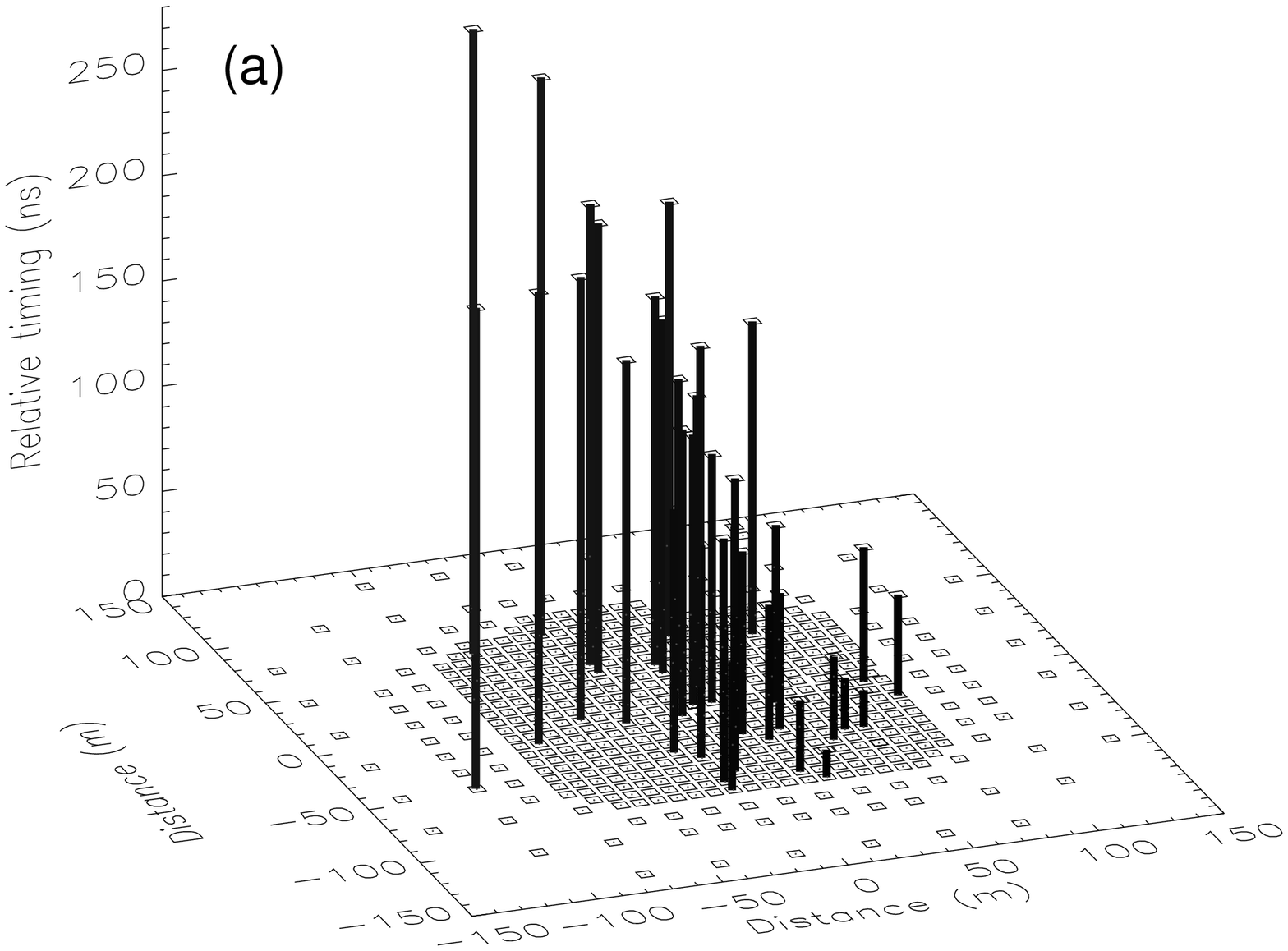}
\end{center}
\end{minipage}
\begin{minipage}{8.cm}
\begin{center}
\plotone{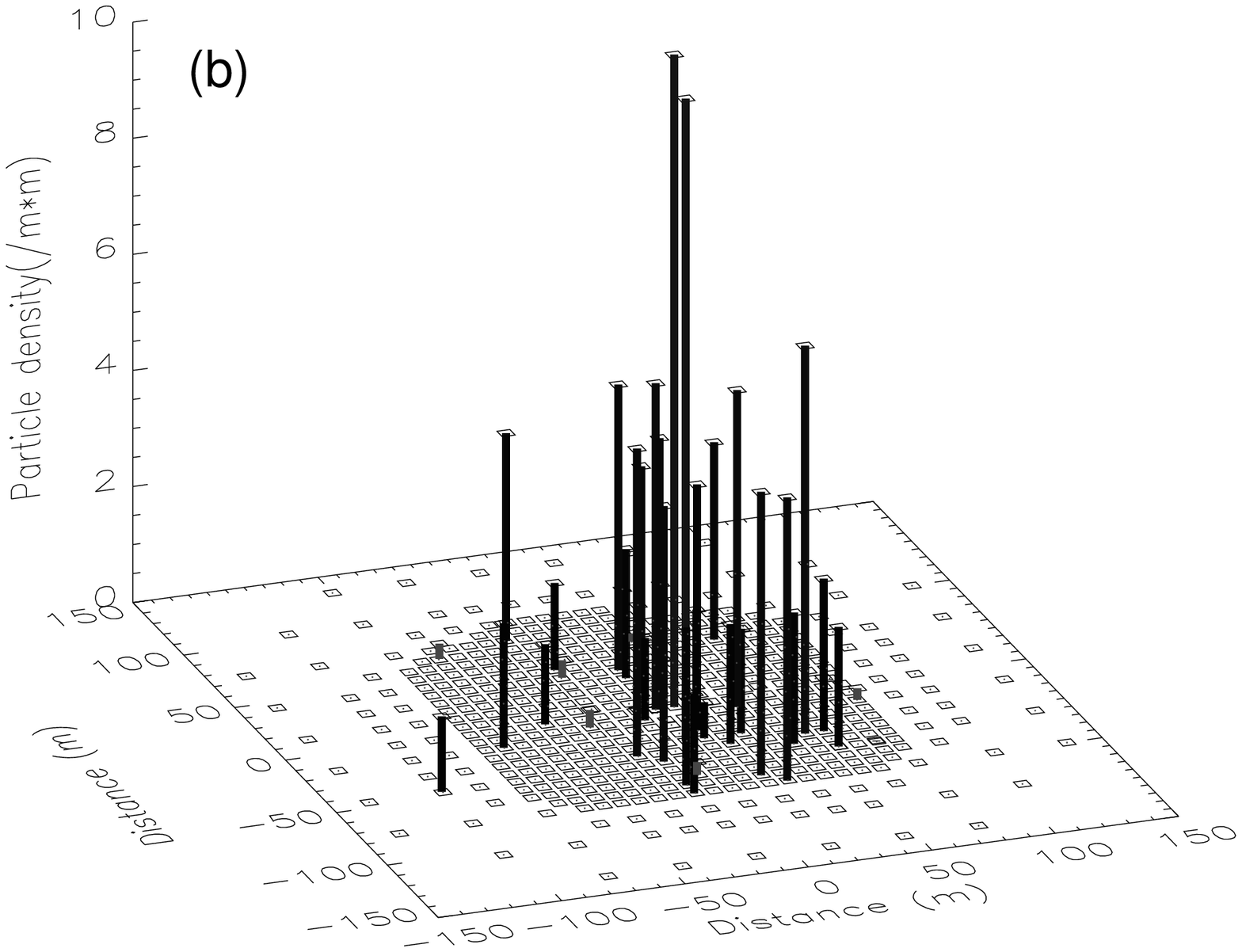}
\end{center}
\end{minipage}
\caption{An example of the map of (a) arrival time of shower particles,
(b) particle density,  obtained by Tibet-III array. 
}
\label{fig:6}
\end{figure}

\section{Analysis}

\subsection{Reconstruction of air-showers}

An example of the shower profile obtained by Tibet-III array is shown in
Fig.~\ref{fig:6}(a) and Fig.~\ref{fig:6}(b) which represent 
the map of arrival time and particle density of shower particles, respectively.
Although the Tibet array has quite low energy threshold of a few TeV for the
purpose of the celestial gamma ray observation, the detection
efficiency for the nuclear component including iron nuclei is not
sufficient at low energy region.  Additional event selection condition 
is required for the unbiased detection of all particles and for 
the capability of the lateral density fitting.
The following condition was applied on the selection of the events
for the all-particle spectrum analysis. 

\begin{equation}
\label{criterion1}
N_{D}\geq 10 \mbox{ with } n_{p} \geq 5 \mbox{ ,}
\end{equation}
where $N_{D}$ expresses the number of detectors hit and
$n_p$ the number of particles per a detector.
This condition satisfies the requirements for unbiased analysis
in the energy range above 100 TeV as described below.

%
% Fig.7
%
\begin{figure}
\epsscale{.70}
\plotone{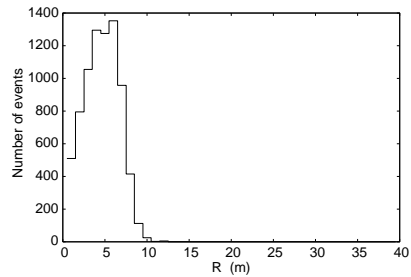}
\caption{The distribution of core-position error. 
The mean error of core position can be estimated as 5 m.
Shower selection criteria : $E_0$ $\geq$ 100 TeV, 
sec($\theta$) $\leq$ 1.1 and the core position located at the inner 135 m 
$\times$ 135 m of the array. 
}
\label{fig:7}
\end{figure}

\subsubsection{
Determination of the core position.}

The core position of each air-shower ($X_{core}$, $Y_{core}$) 
is estimated by using the following equation :
\begin{equation}
\label{core}
(X_{core}, Y_{core}) = (\frac{\Sigma {\rho_i}^{w}{x_i}}{\Sigma {\rho_i}^{w}}, \frac{\Sigma {\rho_i}^{w}{y_i}}{\Sigma {\rho_i}^{w}})
\mbox{ ,}
\end{equation}
where $\rho_i$ is the particle density at the $i$-th detector and the weight 
$w$ is an energy dependent parameter varying between 0.8 and 2.0. 
It is confirmed that the mean error of the core position
can be estimated as 5 m by reconstructing MC events (see Fig.~\ref{fig:7}).
Lower energy event selection condition than eq.~(\ref{criterion1}) leads to
worse core resolution which makes lateral density fitting difficult.

%
% Fig.8
%
\begin{figure}
\epsscale{.70}
\plotone{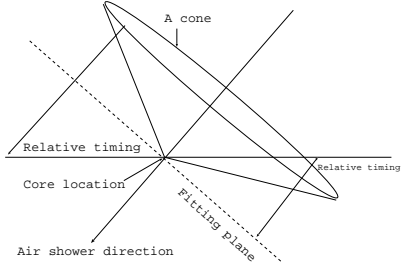}
\caption{Determination of air-shower direction.
}
\label{fig:8}
\end{figure}

\subsubsection{Determination of the arrival direction.}
The arrival direction of the air shower is estimated using the 
time signal measured by the 761 FT (fast-timing) counters. 
The shape of the shower front is assumed to be a reverse-conic type 
as shown in Fig.~\ref{fig:8}.
Direction cosine of the shower axis is determined by using a method of least squares
in which the difference is minimized between the arrival time signals of 
each detector and the expected values on the assumed cone with 
given direction cosine. 

%
% Fig.9
%
\begin{figure}
\epsscale{.70}
\plotone{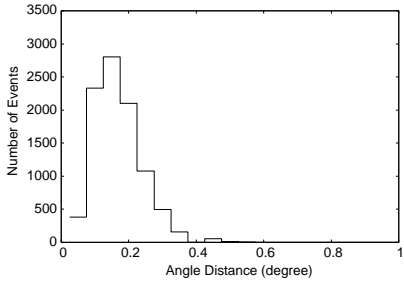}
\caption{
The distribution of opening angle between true and estimated arrival
directions. The mean error of air-shower direction can be estimated
as 0.2$^{\circ}$.
Shower selection criteria : $E_0$ $\geq$ 100 TeV,
sec($\theta$) $\leq$ 1.1 and the core position located at the inner 135 m
$\times$ 135 m of the array.
}
\label{fig:9}
\end{figure}

An experimental check of the angular resolution by this method is made by
the observation of the Moon's shadow \cite{Amenomori9} using large statistics
of low energy events ($>$3 TeV) and also a so-called even-odd method, 
which many authors have been using for estimating the angular
resolution \cite{Amenomori4}. 
The reconstruction of the high energy
MC events assures us that the mean error of air-shower direction can be estimated 
as 0.2$^{\circ}$ at energies above 10$^{14}$ eV (see Fig.~\ref{fig:9}).

%
% Fig.10(a) and (b)
%
\begin{figure}[t]
\begin{center}
\begin{minipage}[b]{8.0cm}
\includegraphics*[width=5cm]{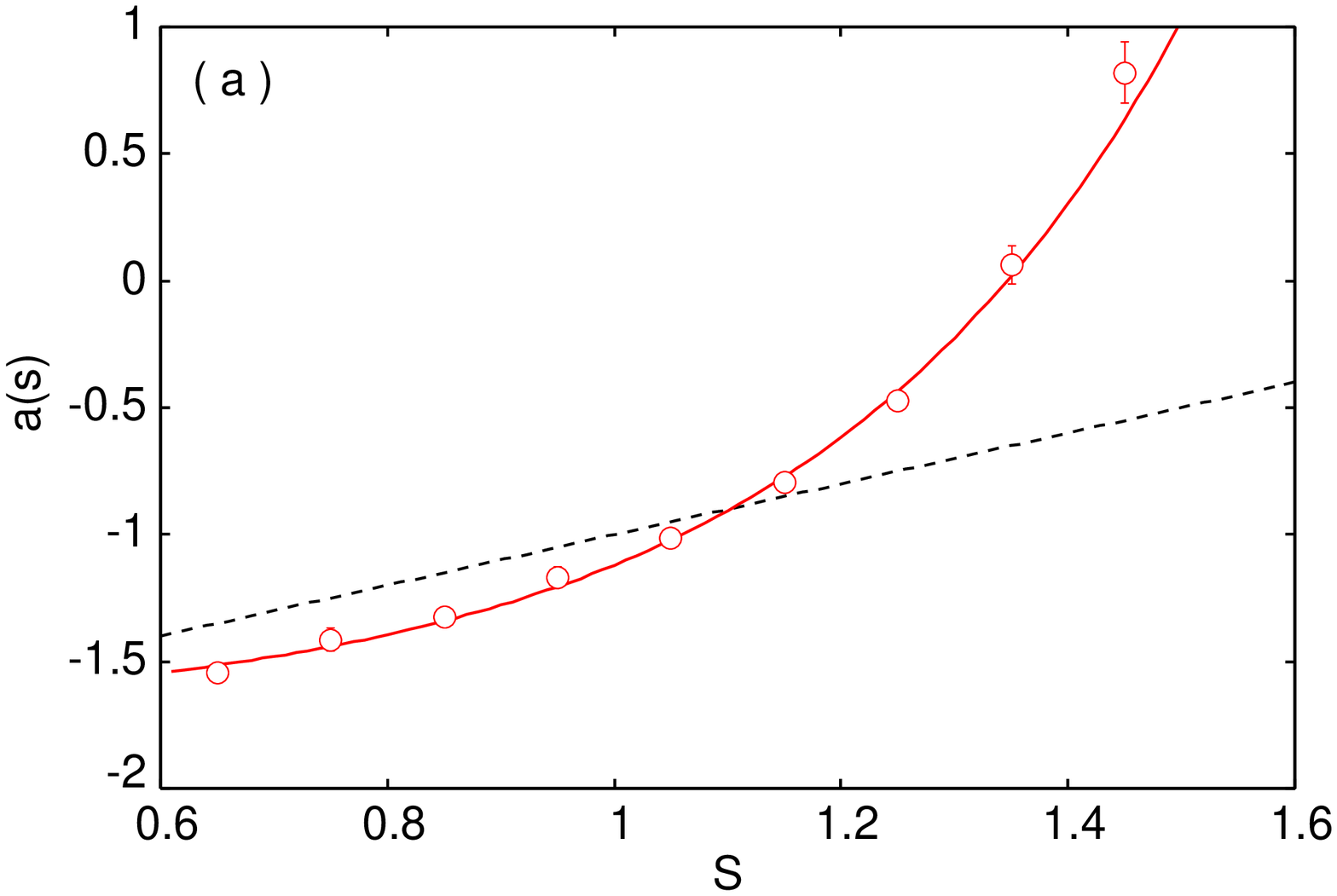}
\end{minipage}
\begin{minipage}[b]{8.0cm}
\includegraphics*[width=5cm]{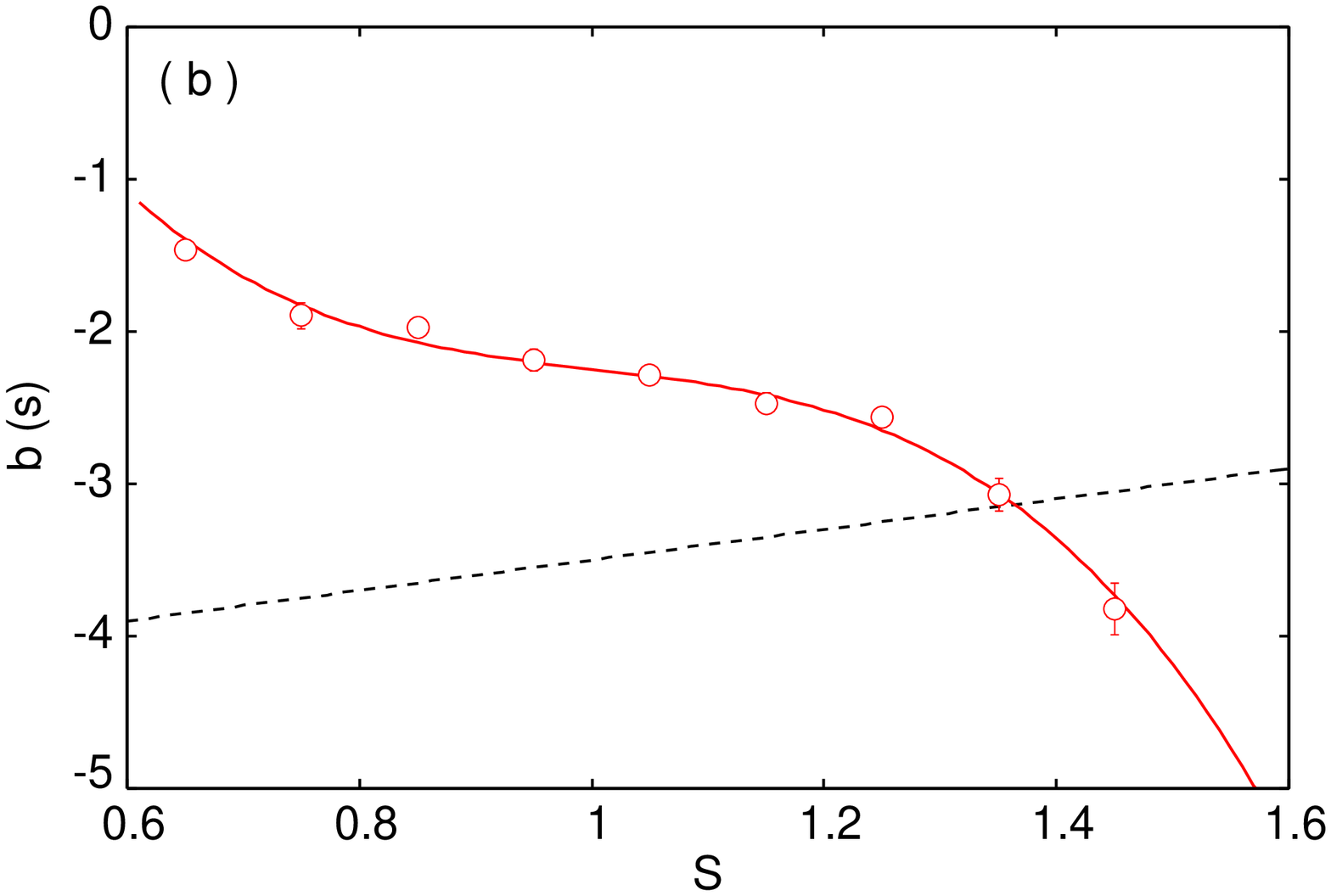}
\end{minipage}
\end{center}
\caption{The numerical values of  $a$ and $b$ are plotted
as a function of $s$, where original definitions
of $a(s)$ and $b(s)$ in NKG function are shown by the dotted lines,
and the open circles denote the averaged MC data using QGSJET01c+HD
model which are
fitted by empirical formulae shown by red lines.  See the text.
}
\label{fig:10}
\end{figure}

\subsubsection{Estimation of the shower size.}
\label{ModNKG}

%
% Fig.11
%
\begin{figure}
\epsscale{.70}
\plotone{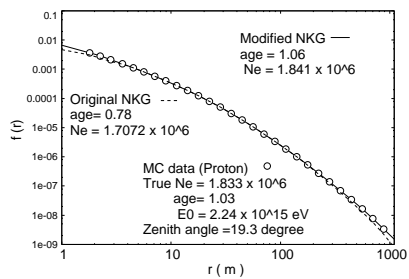}
\caption{
Lateral density distribution of the charged particle obtained
with use of the carpet simulation. The shower size Ne is better reproduced 
by our modified NKG function. 
}
\label{fig:11}
\end{figure}

The lateral density distribution is corrected to the inclined plane
perpendicular to the shower axis and used for the shower size estimation.
In this work, the determination of the lateral distribution 
function of shower particles is very important, since the 
total number of charged particles in each event is estimated 
by fitting this function to the experimental data. 
Using the Monte Carlo data obtained under the same 
conditions as the experiment, we found that the following 
modified NKG function can be fitted well to the lateral distribution 
of shower particles under a lead plate of 5 mm thickness: 

\begin{equation}
\label{NKG_func}
f(r,s) = \frac{N_e}{C(s)} ({\frac{r}{{r_{m}}^{'}}})^{a(s)}{(1+{\frac{r}{{r_m}^{'}}})}^{b(s)} /{r_{m}}^{'2}
\end{equation}

\begin{equation}
\label{NKG_func2}
C(s)=2\pi B(a(s) + 2,-b(s)-a(s)-2)
\end{equation}

\noindent
where ${r_{m}}^{'}$ = 30 m,  
and the variable $s$ corresponds to the age parameter, $N_e$ is the total number
of shower particles and $B$ denotes the beta function. 
The original meaning of $r_m$ in NKG formula is a Moliere unit, being 130 m
at Tibet altitude, however, we treat $r_{m}^{'}$ as a unit scale of the
lateral distribution suitable to describe the structure of the air
showers observed by Tibet-III array, whose effective area is 135 m $\times$ 135 m.
The functions $a(s)$ and $b(s)$ 
are determined as follows. In CORSIKA simulation, 
the shower age parameter $s$ is 
calculated at observation level by fitting the number of particles 
to a function for the one dimensional shower development. 
It may be possible to assume that air showers with the same shower age s,
are in almost the same stage of air-shower development in the atmosphere, i.e. 
they show almost the same lateral distribution for shower 
particles irrespective of their primary energies. The lateral distribution 
of the particle density obtained by the simulation with carpet array configuration 
is normalized by the total number of particles which is derived from
the total energy deposit in an infinitely wide scintillator.
These events are then classified according to the stage of
air-shower development using the age parameter
and they are averaged over the classified events. The fitting of the
equation (\ref{NKG_func}) to the averaged MC data is made to
obtain the numerical values $a$ and $b$. Thus, we can obtain the behavior of
$a$ and $b$ as a function of $s$ as shown in Fig.~\ref{fig:10}(a) and 
Fig.~\ref{fig:10}(b), where original definitions of $a(s)$ and $b(s)$ in NKG function 
are shown by dotted lines. Although our result shows different dependences
of $a$ and $b$ on $s$ from the original NKG function, it is confirmed that the lateral 
distribution of the shower particles is better reproduced by our 
formula (see Fig.~\ref{fig:11}).
This expression is valid in the range of s = 0.6 $\sim$ 1.6, 
$\sec\theta < $ 1.1 and r = 5 $\sim$ 3000 m. 
Two interaction models of QGSJET01c and SIBYLL2.1 
and four primary composition models of pure proton, pure iron, 
HD and PD are used independently to determine
the functions $a(s)$ and $b(s)$ and used in the analysis described below.
Other details are described in Ref.\cite{Amenomori6}.

%
% Fig.12(a),(b)
%
\begin{figure}
\begin{minipage}{8.cm}
\begin{center}
\plotone{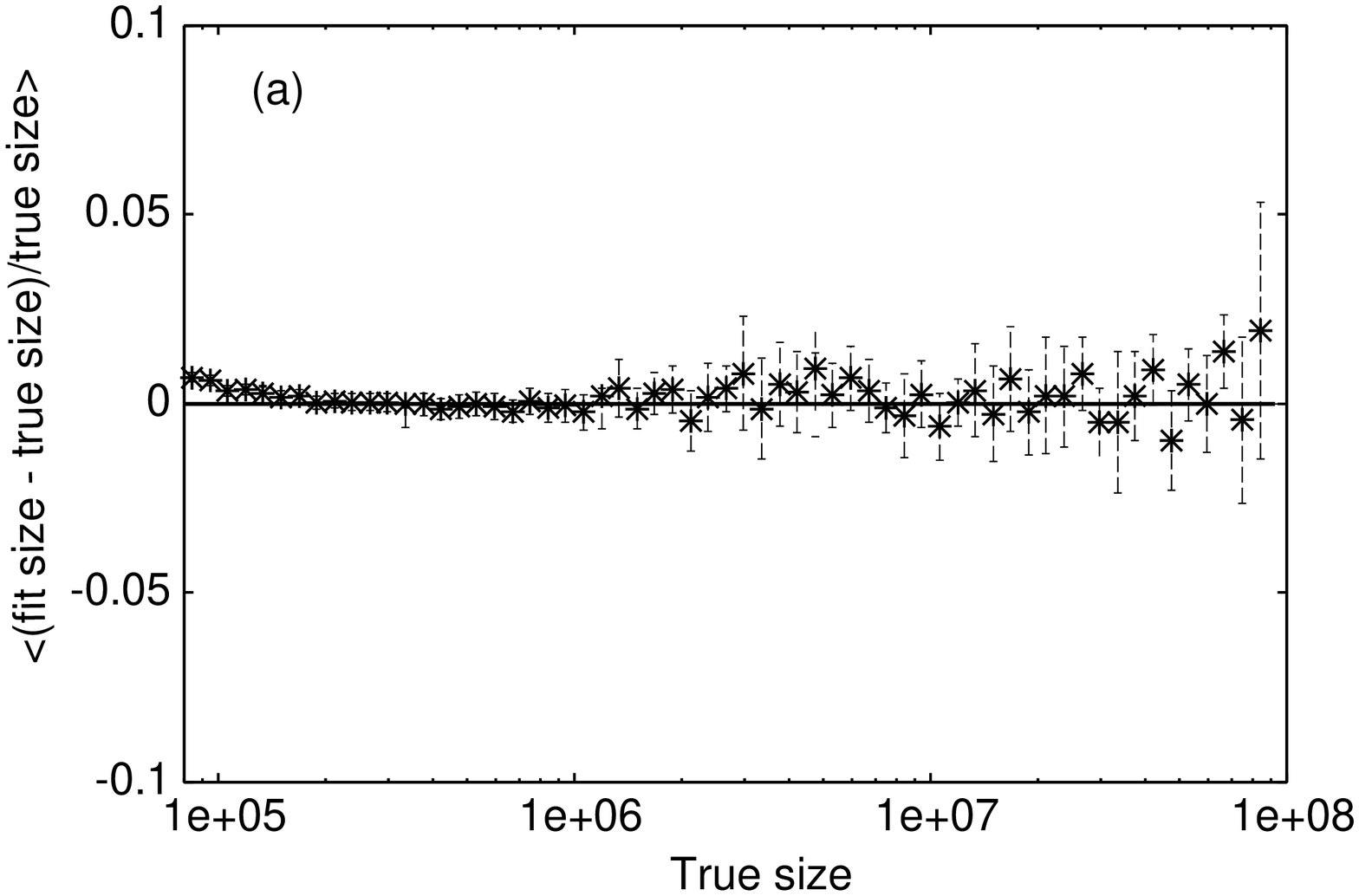}
\end{center}
\end{minipage}
\begin{minipage}{8.cm}
\begin{center}
\plotone{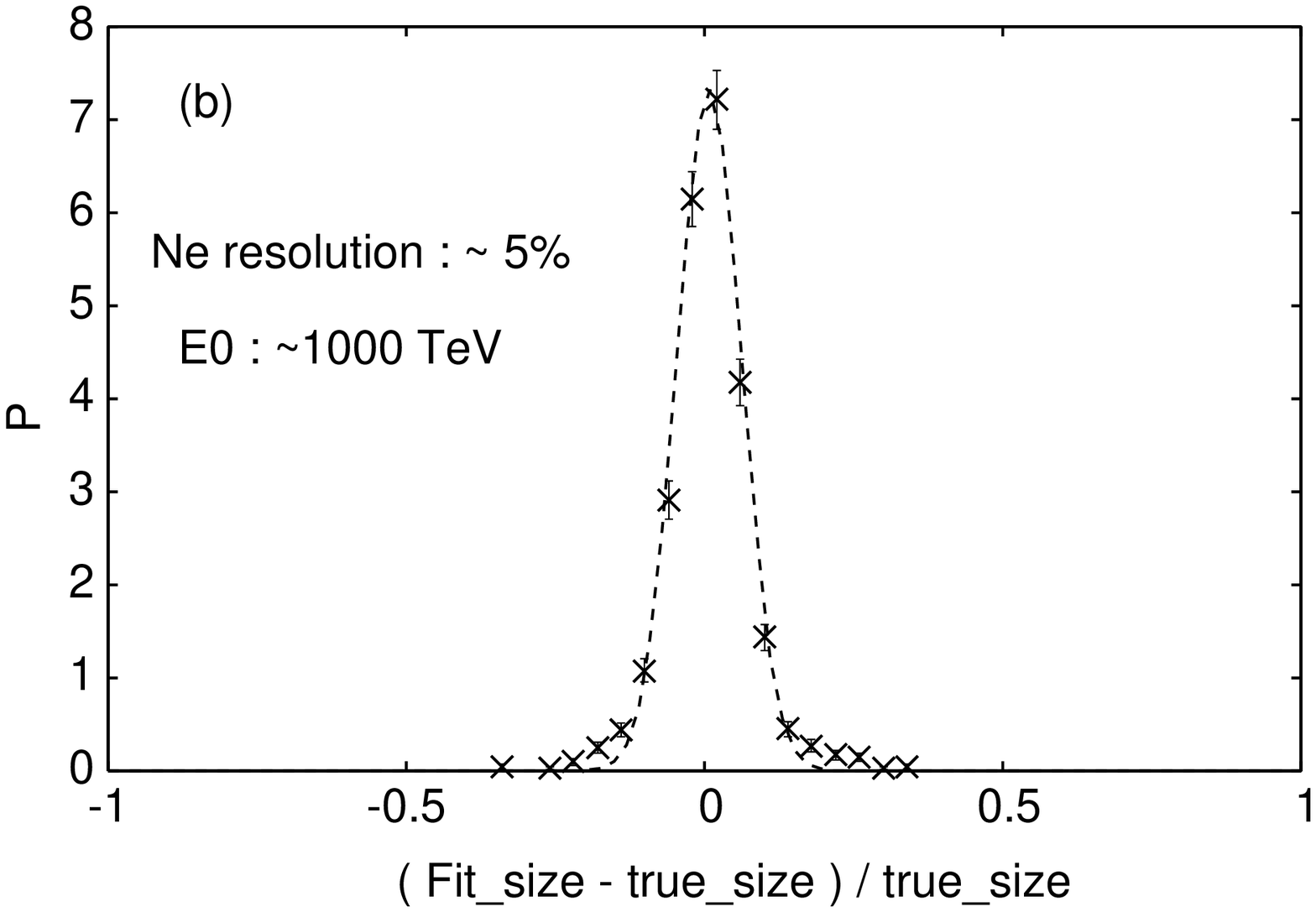}
\end{center}
\end{minipage}
\caption{(a) The correlation between the true shower size and the estimated shower size 
(fit size). (b) The shower size resolution is estimated to be 5\% around the primary 
energy of 1000 TeV based on the QGSJET+HD model. 
Shower selection criteria : $E_0$ $\geq$ 100 TeV, 
sec($\theta$) $\leq$ 1.1 and the core position located at the inner 135 m $\times$ 135 m of the array. 
}
\label{fig:12}
\end{figure}
 
Based on the Monte Carlo simulation, the correlation between the true shower 
size (true size) and the estimated shower size (fit size) is demonstrated in 
Fig.~\ref{fig:12}(a) and Fig.~\ref{fig:12}(b). Here, the true shower size means 
the number of particles calculated for a carpet array while estimated shower size is 
for the real Tibet-III array using the modified NKG function mentioned above. 
As seen in these two figures, a good correlation is obtained 
between the true shower size and the estimated shower size. 
The systematic deviation of less than 1 \% seen around 100 TeV 
of Fig.~\ref{fig:12}(a)
shows that we need finer tuning of modified NKG function at low energies and
this error is finally corrected.
The shower size 
is well reproduced with standard deviation of 5\% around the primary energy 
of 1000 TeV with 1.0 $<$ sec($\theta$) $\leq$ 1.1 based on the QGSJET+HD model.
The shower size resolutions estimated are summarized for 
the events with different simulation model combinations in Table~\ref{tab:01}.

%
% TABLE I.
%
\begin{table}[t]
\begin{center}
\caption{The shower-size resolution  are summarized for the events
induced by primary particles of $E_0$ $\simeq$ 100 TeV
and $E_0$ $\simeq$ 1000 TeV
with sec($\theta$) $\leq$ 1.1 and the core position located at inner 135 m
$\times$ 135 m of the array,
based on the QGSJET+HD, QGSJET+PD, SIBYLL+HD and SIBYLL+PD model.}
\begin{tabular}{l|cccc}
\tableline\tableline
 & & E0 = 100 TeV & & \\
\tableline
Zenith angle &QGS.&QGS.&SIB.&SIB.\\
(sec($\theta$))&+HD&+PD&+HD&+PD\\
\tableline
1.0 - 1.1&9.0\%&9.1\%&9.2\%&9.5\%\\
\tableline
\tableline
 & & E0 = 1000 TeV & & \\
\tableline
Zenith angle&QGS.&QGS.&SIB.&SIB.\\
(sec($\theta$))&+HD&+PD&+HD&+PD \\
\tableline
1.0 - 1.1 &5.0\%&5.8\%&5.2\%&5.9\% \\
\tableline
\tableline
\end{tabular}
\label{tab:01}
\end{center}
\end{table}

\subsection{Data selection}
\label{data-selection}

The following event-selection criteria are
adopted in the present analysis.

\noindent 
1) More than 10 detectors should detect a signal of more than five particles 
per detector, as mentioned in eq.(\ref{criterion1}).

\noindent 
2) In order to minimize the primary mass dependence on the air-shower size 
at Yangbajing altitude, the zenith angle ($\theta$) of the arrival direction
of air showers
should be smaller than 25$^{\circ}$, or $\sec\theta$ $\leq$ 1.1. 

\begin{figure}
\epsscale{.70}
\plotone{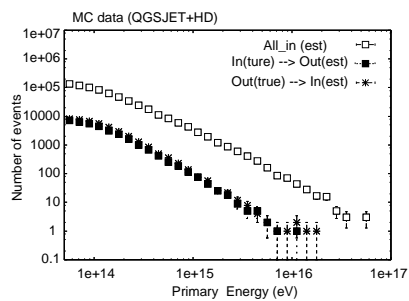}
\caption{
Out-In events occupy about 10\% of all selected events.
The energy spectrum of Out-In and In-Out events are almost the same,
being the effect of the difference between them to the
all selected events less than 2\%  at each energy bin.
}
\label{fig:13}
\end{figure}

\noindent 
3) The rejection of the events falling outside the effective area 
of the array is made by
estimating the core position using eq. (\ref{core}) with high weight
of $w=8.0$ for the particle density. 
Then, we imposed that the core position should be inside the 
innermost 135 m $\times$ 135 m area (18225 $m^2$). 
This area is chosen with the use of 
MC events, so that the following two cases are just canceling out each other, 
namely the number of events originally inside of this area but falling 
outside after event reconstruction equals to the number of events 
in the opposite case. It is confirmed by simulations that Out-In events 
occupy about 10\% of all selected events (see Fig.~\ref{fig:13}), 
and the energy spectrum of Out-In and In-Out events are almost the same, 
hence the effect of the difference between them to the all selected events 
is less than 2\%  at each energy bin.

%
% Fig.13
%

\subsection{Trigger efficiency}
\begin{figure}
\begin{minipage}{8.5cm}
\begin{center}
\plotone{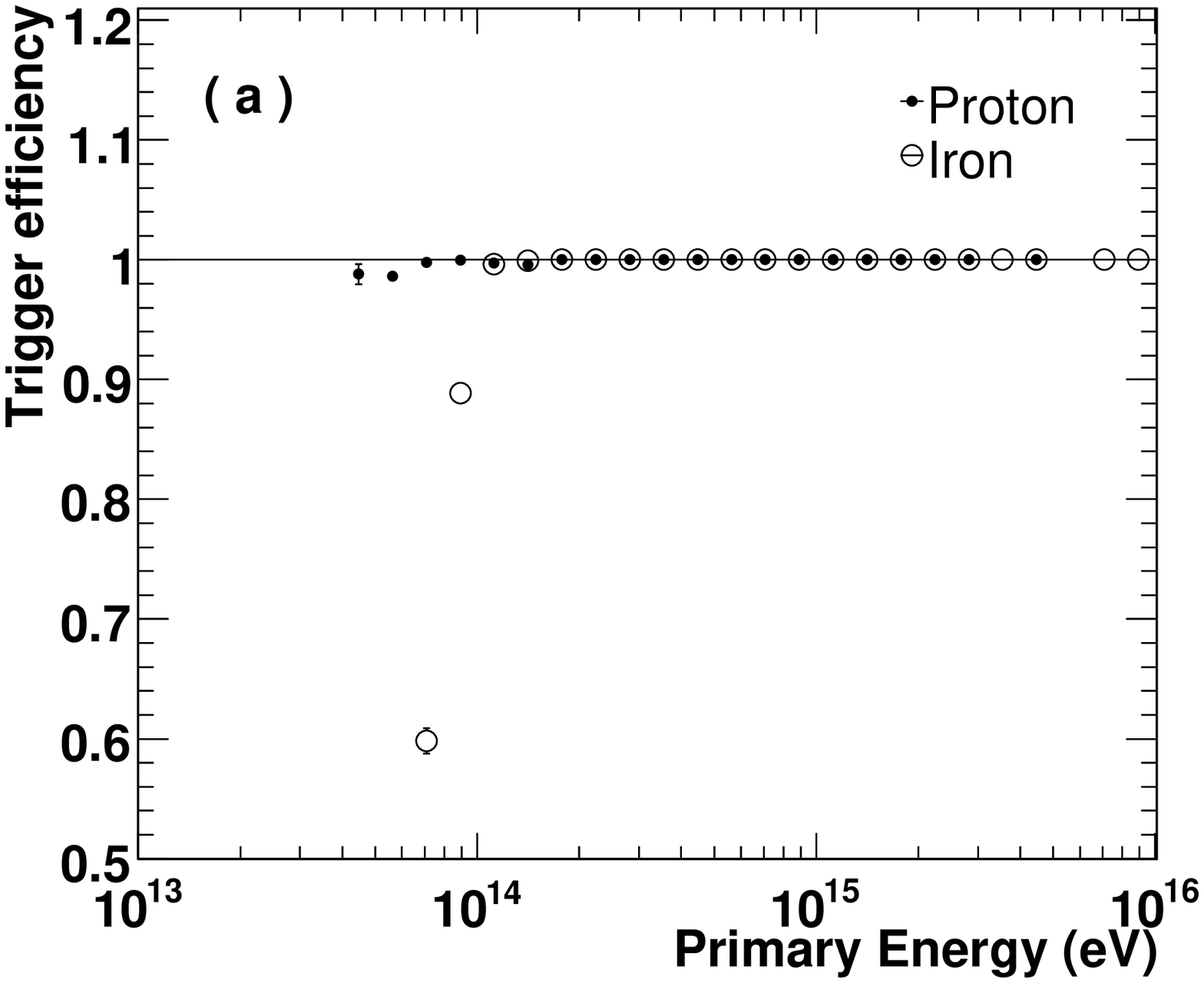}
\end{center}
\end{minipage}
\begin{minipage}{8.5cm}
\begin{center}
\plotone{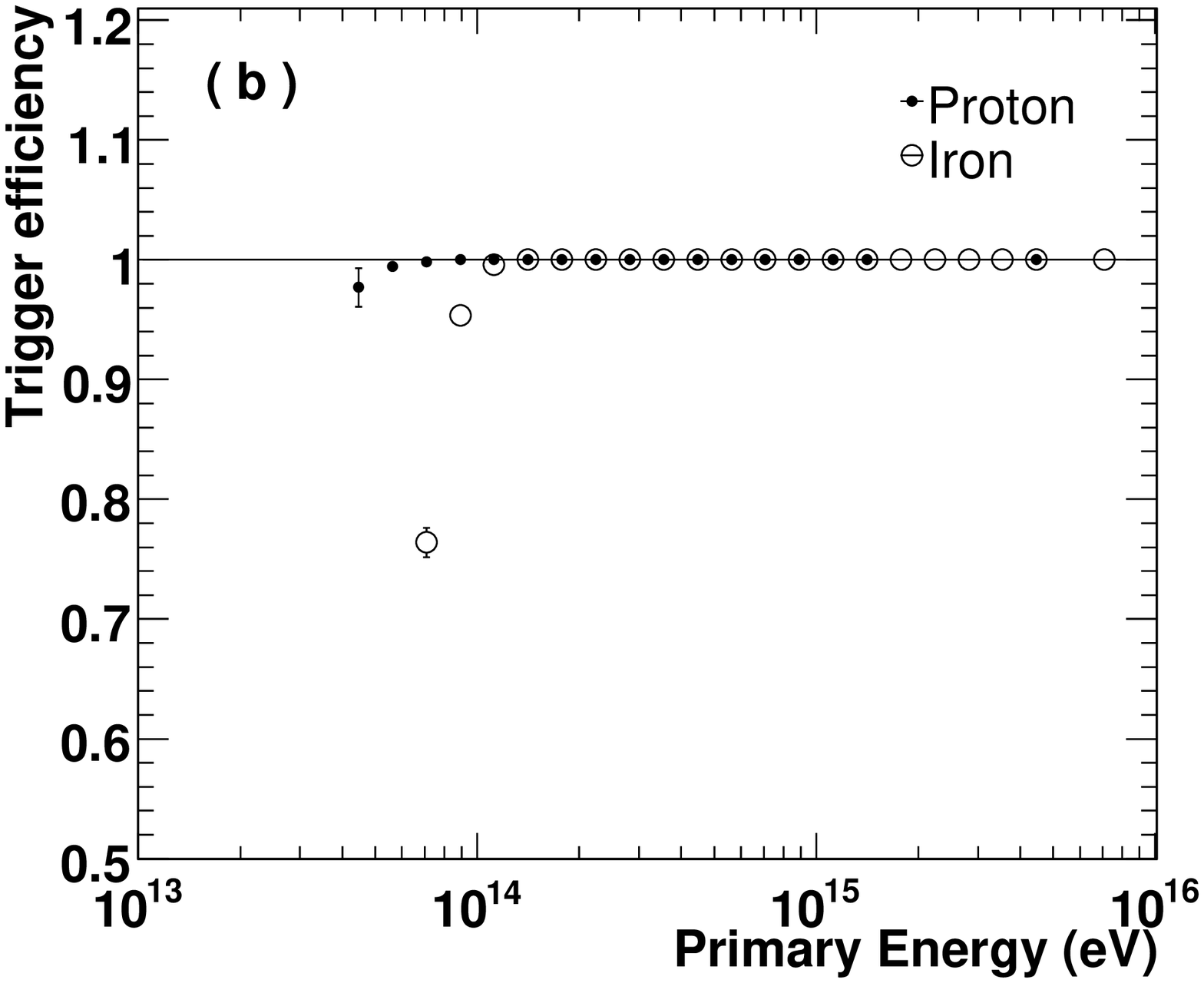}
\end{center}
\end{minipage}
\caption{The trigger efficiency of air showers. In the case of $N_{D}$ $\geq$ 10,
$n_{p}$ $\geq$ 5 and sec($\theta$) $\leq$ 1.1 with the core position located
at the inner
135 m $\times$ 135 m of the array, the air showers induced by protons
or irons with $E_0$ $\geq$ 100 TeV can be fully detected without any bias.
(a) QGSJET and (b) SIBYLL.}
\label{fig:14}
\end{figure}

Our simulations confirmed that the air showers induced by primary 
particles with $E_0$ $\geq$ 100 TeV 
can be fully detected without any bias under the above mentioned criteria 
as shown in Fig.~\ref{fig:14}.
The total effective area S $\times$ $\Omega$ is then calculated to be 
10410 m$^{2}\cdot$sr for all primary particles with $E_0$ $\geq$ 100 TeV
using inner area of 135 m $\times$ 135 m and solid angle for 
sec($\theta$) $\leq$ 1.1. For the calculation of the absolute intensity,
the inclination effect due to a flat surface detector is taken
into account by correcting the density of the observed events
into that for a plane perpendicular to the shower axis.
For the operation period from 2000 November through 2004 October, 
the effective live time $T$ is calculated as 805.17 days. 
The total number of air showers selected under the above conditions 
is 5.5 $\times$ $10^{7}$ events after the inclination correction.

%
% Fig.14(a),(b)
%

\section{Results and Discussions}

\subsection{The model dependence of the size spectrum}

%
% Fig.15
%
\begin{figure}
\epsscale{.80}
\plotone{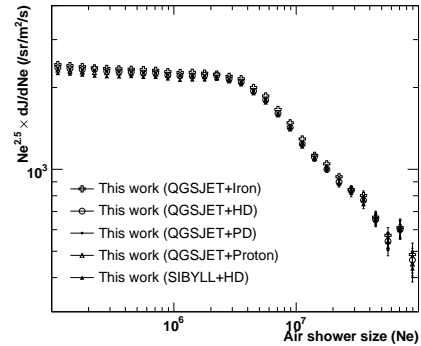}
\caption{The model dependence of the size spectrum of 
nearly vertical air showers (sec($\theta$) $\leq$ 1.1).}
\label{fig:15}
\end{figure}

As a first step in checking the model dependence, the difference
of the size spectra derived by the lateral fitting is examined
using the structure functions based on five models 
(QGSJET interaction model with four
primary composition models and SIBYLL+HD model).
Shown in Fig.~\ref{fig:15} is the model dependence of 
the air-shower size spectrum of nearly vertical air showers. 
It is seen that the model dependence of the air-shower size 
is small (less than 5\% in absolute intensity on the primary composition model, 
and also less than 5\% on the hadronic interaction models).

\subsection{All-particle energy spectrum}
\subsubsection{Determination of the primary energy}
%
% Fig.16
%
\begin{figure}
\epsscale{.80}
\plotone{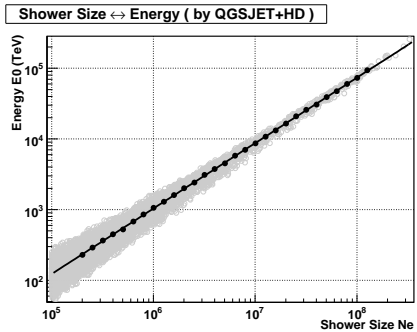}
\caption{Scatter plots of the primary energy $E_0$ and the estimated shower
size $N_e$ based on the QGSJET+HD model.
}
\label{fig:16}
\end{figure}

In Fig.~\ref{fig:16} we show the scatter plots of the primary energy $E_0$,
and the estimated shower size $N_e$ based on the QGSJET+HD model.
%
% Fig.17
%
\begin{figure}
\epsscale{.80}
\plotone{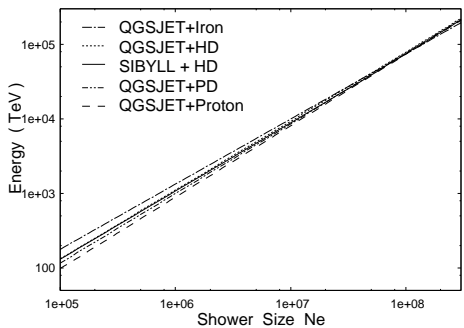}
\caption{The correlations between the estimated shower size $N_e$
and the primary energy $E_0$ for given models.
}

\label{fig:17}
\end{figure}

%
% Fig.18
%
\begin{figure}
\epsscale{.80}
\plotone{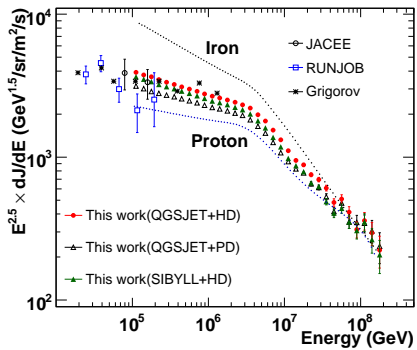}
\caption{The differential energy spectra of all particles obtained 
by the present work using 5 models
and they are compared with direct observations.
JACEE \cite{Asakimori},RUNJOB \cite{Apanasenko},
Grigorov \cite{Grigorov}
}
\label{fig:18}
\end{figure}

Fig.~\ref{fig:17} shows the correlations between the estimated shower size $N_e$ 
and the primary energy $E_0$ for QGSJET + 4 primary models
and SIBYLL+HD model. In this figure,
the results of pure composition models are also shown just to help the
understanding of the composition dependence in the energy determination.
Using these pure composition model will result in remarkably different
primary energy spectrum as shown later.
The correlation between the estimated shower size $N_e$  
and the primary energy $E_0$ for $\sec\theta$ $\leq$ 1.1
can be well fitted using a following conversion function
for each interaction models and composition models.

\begin{equation}
\label{conver}
E_0 = a \times {(\frac{N_e}{1000})}^{b} \mbox{\hspace{2cm}  [TeV]},
\end{equation}
\noindent
where the numerical values of $a$ and $b$ in eq.(\ref{conver}) are summarized in
Table ~\ref{tab:01a}. These approximations are valid
for nearly vertical air showers (sec($\theta$) $\leq$ 1.1 )
at Yangbajing altitude.

As seen in Fig.~\ref{fig:17}, the difference of the conversion factor
between two mixed composition models of HD and PD is not significant 
and almost disappears above a few times 1000 TeV in spite of the
large difference of the fractional abundances of chemical components.
From the comparison between QGSJET+HD and SIBYLL+HD,
one can see that the dependence on the interaction model is very small.

It is also noted that the energy resolutions in different interaction models and two
mixed composition models are very close to each other.
The differences are estimated as 
36\% and 17\% at energies around 200 TeV and 2000 TeV, respectively,
in the case of QGSJET+HD model. 
Corresponding values for SIBYLL+HD model are 38\% and 19\% and
that for QGSJET+PD model are 39\% and 19\%.
It is commonly understood that there is an energy dependent overestimation problem
of the flux due to the steep power index of the cosmic ray energy
spectrum when the error of the estimated energy increases with decreasing 
primary energy.
It may be, however, worthwhile to note here that this effect is already 
included in our method of the 
energy determination using the conversion function,
because the primary energy at a given shower size
is determined including the contribution of all possible primary energies
which leads to a smaller value than the case of without fluctuation
reflecting the larger population of low energy primaries than high energies.
To avoid methodical systematic error,
the reproducibility of the primary flux was carefully examined using MC events
and no significant deviation was found between MC input spectrum
and the reconstructed spectrum.

%
% Fig.19
%
\begin{figure}
\epsscale{.80}
\plotone{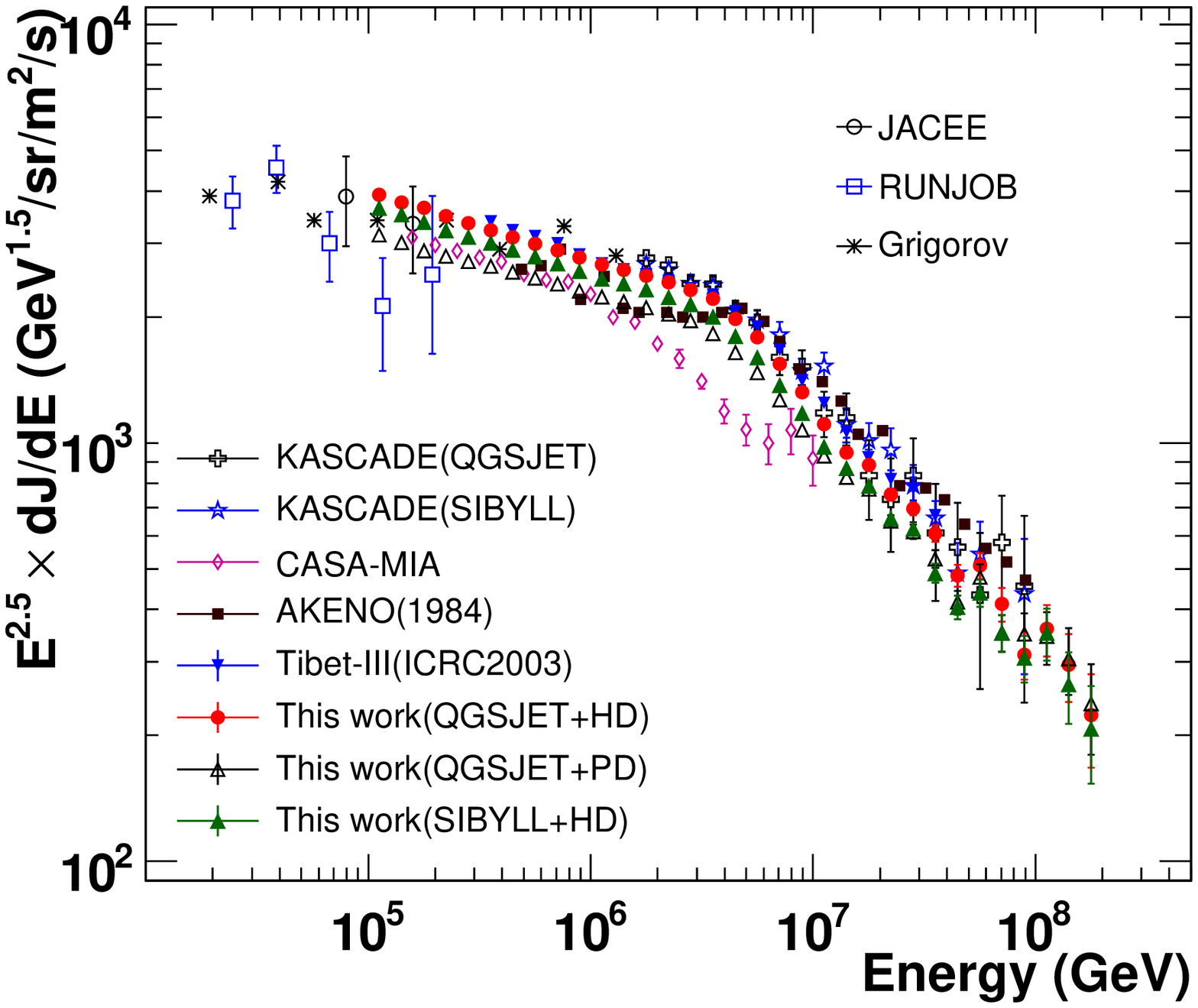}
\caption{The differential energy spectra of all particles obtained 
by the present work using mixed composition models 
compared with other experiments. 
JACEE \cite{Asakimori},RUNJOB \cite{Apanasenko},
Grigorov \cite{Grigorov}, KASCADE \cite{Antoni},
CASA-MIA \cite{Glasmacher}, AKENO(1984) \cite{Nagano1}, 
Tibet-III(ICRC2003) \cite{Amenomori8}.
}
\label{fig:19}
\end{figure}

\subsubsection{Energy spectrum of all-particles and the knee parameters}

%
% Fig.20
%
\begin{figure}
\epsscale{.80}
\plotone{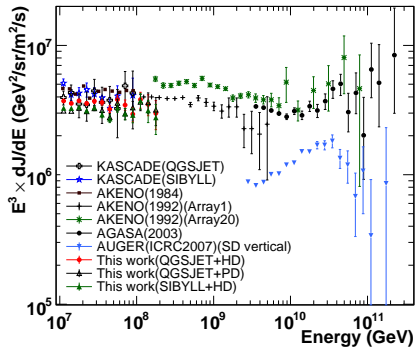}
\caption{The differential energy spectra of all particles obtained 
by the present work above the knee 
compared with other experiments at the highest energy range.
KASCADE \cite{Antoni}, AKENO(1984) \cite{Nagano1},
AKENO(1992) \cite{Nagano2}, AGASA(2003) \cite{AGASA}, 
AUGER(ICRC2007)(SD vertical) \citep{Yamamoto,Roth}
}
\label{fig:20}
\end{figure}

The all-particle energy spectrum of primary cosmic rays in a wide range 
over 3 decades between 1 $\times$ $10^{14}$ eV  and 1 $\times$ $10^{17}$ eV 
is shown in Fig.~\ref{fig:18} for 5 models.
In this figure, it is important to note that the position of the knee is clearly seen at
the energy around 4 PeV irrespective of the model used.

The model dependence of the conversion function shown in 
Fig.~\ref{fig:17} may lead to different shapes of the all-particle
energy spectrum. The model dependence on the primary
composition can be checked by comparing the results of
QGSJET+HD and QGSJET+PD. Although the composition is fairly different between the HD and
PD model at energies above $10^{15}$ eV, the difference of the absolute
intensity is 20\% at most between the two models and decreases with increasing
primary energy. As mentioned above, it may be helpful to show the composition
dependence by the pure component assumption.
The difference between pure proton and pure iron primary model becomes
large at the lower energy region of the range shown in this figure, where
the difference of the intensities between two models
exceeds a factor of 3 at 10$^{14}$ eV, although 
these results do not belong to the argument for the reality at least
in the energy range lower than a few times 10$^{14}$ eV.
It should be noted that the
mixed composition models give more realistic extension of the
direct observations and the composition model dependence almost
disappears above 10$^{16}$ eV. This is the remarkable
characteristics of the Tibet experiment.

The interaction model dependence is seen by
comparing the results of QGSJET+HD and SIBYLL+HD. This comparison shows that
the shapes of the
spectrum from both interaction models are almost the same and the difference in
the absolute intensity is within 10\%. 
In Fig.~\ref{fig:19}, we show 
the results using the mixed composition model
in comparison with other works including our previous work presented in ICRC2003. 
Our results seem to lead to the data of direct observation smoothly 
in the low energy side. Fig.~\ref{fig:20} shows the higher energy part 
above the knee to compare with the surface array experiments done 
at the highest energy region. It is very interesting to see how the smooth extension of our
 spectrum above $10^{16}$ eV is connected with the data at the highest energy region. 

The intensities of all-particle energy spectra 
measured by the Tibet-III array are also posted in Table~\ref{tab:03}
and the summary of 
the measured all-particle energy spectrum and knee parameters 
are listed in Table~\ref{tab:02}, where $\gamma_1$ is the best fitted-index 
for the energy range 100 TeV $<$ $E_0$ $<$ 1 PeV, and $\gamma_2$ is for the energy above 4 PeV.

%
% TABLE II.
%
\begin{table}[t]
\begin{center}
\caption{The summary of the knee parameters. The symbol 
$\gamma_1$ is the best fitted-index for the energy range 
100 TeV $<$ $E_0$ $<$ 1 PeV,
and $\gamma_2$ is for the energy above 4 PeV. 
}
\begin{tabular}{l|c|c}
\tableline\tableline
Model & knee position & Index of \\
 & (PeV) & spectrum \\
\tableline
QGSJET& (4.4 $\pm$ 0.1) & $\gamma_1$= -2.81 $\pm$ 0.01 \\
+Iron&  & $\gamma_2$= -3.21 $\pm$ 0.01 \\
\tableline
QGSJET& (4.0 $\pm$ 0.1) & $\gamma_1$= -2.67 $\pm$ 0.01 \\
+HD&  & $\gamma_2$= -3.10 $\pm$ 0.01 \\
\tableline
QGSJET& (3.8 $\pm$ 0.1) & $\gamma_1$= -2.65 $\pm$ 0.01 \\
+PD &  & $\gamma_2$= -3.08 $\pm$ 0.01 \\
\tableline
SIBYLL & (4.0 $\pm$ 0.1) & $\gamma_1$= -2.67 $\pm$ 0.01  \\
+HD &  & $\gamma_2$= -3.12 $\pm$ 0.01  \\
\tableline
QGSJET& (3.4 $\pm$ 0.1) & $\gamma_1$= -2.60 $\pm$ 0.01 \\
+Proton &  & $\gamma_2$= -3.03 $\pm$ 0.01 \\
\tableline
\tableline
\end{tabular}
\label{tab:02}
\end{center}
\end{table}

\section{Summary}
\vspace{-3mm}
We have analyzed the air shower data set collected in the period from 2000 November through 
2004 October with the Tibet-III air-shower array, using the new simulation code
CORSIKA, and obtained the all-particle  energy spectrum 
of primary cosmic rays in a wide energy range over 3 decades between 
$10^{14}$ eV and $10^{17}$ eV. The knee of the primary spectrum 
is clearly observed and its position is located at the energy around 4 PeV.

The advantage of the Tibet experiment at high altitude is first that the
primary energy for the unbiased detection of air showers is sufficiently low 
for the purpose of the measurement around the knee, and second that the
energy determination is insensitive to the number of muons which is dependent on
hadronic interaction model and the chemical composition. 
In order to quantitatively confirm these characteristics,
the model dependence on the primary
chemical composition was estimated in terms of the two mixed 
chemical composition models of HD (heavy dominant) and PD (proton dominant),
which are extrapolated from the direct observations with different scheme
of the fractional contents of the individual elements,
together with the extreme cases of pure proton and pure iron.
The interaction model dependence was discussed using QGSJET01c and
SIBYLL2.1 interaction models as they are widely used by other works.
It was shown that the air showers induced by primary energies above 100 TeV
are fully detected for all kind of primary particles.
The systematic errors due to the above mentioned model dependences
are shown to be within a few tens \% in the energy range below the knee,
i.e., 20\% in chemical
composition models between HD and PD, and 10\% in interaction
models between QGSJET01c and SIBYLL2.1. The model dependence is
decreasing with increasing primary energy, 
and it almost disappears above 10$^{16}$ eV.
Although these estimates are limited for the chosen models, 
other choice of more adequate models, if any, will not change the
result of this work drastically because of the weak dependence on the
model used.
The uncertainty due to the interaction model will  even decrease after
the measurement of the CMS forward region by the forthcoming 
experiment LHCf~\citep{LHCf,Sako2007}.
This is the highest-statistical and the best 
systematics-controlled measurement covering the widest energy range around the 
knee energy region. 

As discussed above, the main uncertainty left in the all-particle spectrum
is related to the chemical composition.
This work will be extended to the analysis of the air showers with
large arrival zenith angle which can show another features of the air-shower
development in the atmosphere and provides the information about the
chemical composition (to be published elsewhere).
As mentioned before,
we have already reported a paper on the proton and helium spectra
around the knee \citep{Amenomori2,Amenomori5} derived from the
hybrid experiment using the air-shower core detector, which is 
sensitive to the showers of light element origin 
like proton or helium by selecting the high energy core.
From the observed steep power index and the
low intensities of proton and helium spectra,
the dominance of the heavy elements were suggested by the hybrid experiment,
however, the statistics were limited due to the high threshold.
In the very near future, we will start a new high-statistics 
hybrid experiment in Tibet \cite{Huang} to clarify the main component of 
cosmic rays at the knee. The core detector will consist of 400 burst detectors
located at the center of the Tibet-III array in a grid with the detector
interval of 3.75m. The burst detectors measure the high energy
electromagnetic cascade of the energy above 30 GeV developed in lead
plate of 3.5 cm thick by shower-core particles.
The new experiment is able to observe the air-shower cores induced by
heavy components around and beyond the knee, where direct measurements are 
inaccessible because of their extremely low fluxes. The first observation
of the iron spectrum in the knee region is expected in this new experiment.

\acknowledgments

The collaborative experiment of the Tibet Air Shower Arrays has been
performed under the auspices of the Ministry of Science and Technology
of China and the Ministry of Foreign Affairs of Japan. This work was
supported in part by Grant-in-Aid for Scientific Research on Priority
Areas from the Ministry of Education, Culture, Sports, Science and
Technology, by Grants-in-Aid for Science Research from the Japan Society 
for the Promotion of Science in Japan, and by the Grants-in-Aid 
from the National Natural Science Foundation of China and the
Chinese Academy of Sciences. The authors thank J. Kota for reading the
manuscript.

%% Use the figure environment and \plotone or \plottwo to include
%% figures and captions in your electronic submission.
%% To embed the sample graphics in
%% the file, uncomment the \plotone, \plottwo, and
%% \includegraphics commands
%%
%% If you need a layout that cannot be achieved with \plotone or
%% \plottwo, you can invoke the graphicx package directly with the
%% \includegraphics command or use \plotfiddle. For more information,
%% please see the tutorial on "Using Electronic Art with AASTeX" in the
%% documentation section at the AASTeX Web site,
%% http://www.journals.uchicago.edu/AAS/AASTeX.
%%
%% The examples below also include sample markup for submission of
%% supplemental electronic materials. As always, be sure to check
%% the instructions to authors for the journal you are submitting to
%% for specific submissions guidelines as they vary from
%% journal to journal.

%% This example uses \plotone to include an EPS file scaled to
%% 80% of its natural size with \epsscale. Its caption
%% has been written to indicate that additional figure parts will be
%% available in the electronic journal.

\clearpage

%
% TABLE Ia.
%
\begin{table}[t]
\begin{center}
\caption{The parameters in the conversion function eq.~(\ref{conver}).
}
\begin{tabular}{l|cc|cc}
\tableline\tableline
Model & E0 $<$ $10^{16}$ eV & & E0 $>$ $10^{16}$ eV & \\
\tableline
  & a & b & a & b \\
\tableline
QGSJET+Proton & 1.195  & 0.959 & 1.195 & 0.959 \\
\tableline
QGSJET+HD & 1.872  & 0.924 & 1.348 & 0.953 \\
\tableline
QGSJET+PD & 1.583  & 0.933 & 1.348 & 0.951 \\
\tableline
SIBYLL +HD & 1.968  & 0.913 & 1.323 & 0.951 \\
\tableline
QGSJET+Iron & 3.915  & 0.851 & 3.915 & 0.851 \\
\tableline
\tableline
\end{tabular}
\label{tab:01a}
\end{center}
\end{table}

\clearpage

%
% TABLE III.
%
\begin{table}[t]
\begin{center}
\caption{The intensity of all-particle energy spectrum measured by Tibet-III array based on the QGSJET+HD, QGSJET+PD 
and SIBYLL+HD models.
}
\begin{tabular}{l|c|c|c}
\tableline
\tableline
Model      &  QGSJET+HD & QGSJET+PD & SIBYLL+HD \\
\tableline
Energy & dJ/dE$\pm$stat.errors & dJ/dE$\pm$stat.errors & dJ/dE$\pm$stat.errors \\
(GeV) & ($m^{-2}{s}^{-1}{sr}^{-1}{GeV}^{-1}$) & ($m^{-2}{s}^{-1}{sr}^{-1}{GeV}^{-1}$) & ($m^{-2}{s}^{-1}{sr}^{-1}{GeV}^{-1}$) \\
\tableline
1.12 $\times$ $10^{5}$ & (9.300 $\pm$ 0.002)$\times$ $10^{-10}$ & (7.454 $\pm$ 0.002)$\times$ $10^{-10}$ & (8.639 $\pm$ 0.002)$\times$ $10^{-10}$ \\
1.41 $\times$ $10^{5}$ & (5.008 $\pm$ 0.001)$\times$ $10^{-10}$ & (4.013 $\pm$ 0.001)$\times$ $10^{-10}$ & (4.684 $\pm$ 0.001)$\times$ $10^{-10}$ \\
1.78 $\times$ $10^{5}$ & (2.732 $\pm$ 0.001)$\times$ $10^{-10}$ & (2.155 $\pm$ 0.001)$\times$ $10^{-10}$ & (2.529 $\pm$ 0.001)$\times$ $10^{-10}$ \\
2.24 $\times$ $10^{5}$ & (1.470 $\pm$ 0.001)$\times$ $10^{-10}$ & (1.177 $\pm$ 0.001)$\times$ $10^{-10}$ & (1.360 $\pm$ 0.001)$\times$ $10^{-10}$ \\
2.82 $\times$ $10^{5}$ & (7.934 $\pm$ 0.004)$\times$ $10^{-11}$ & (6.437 $\pm$ 0.004)$\times$ $10^{-11}$ & (7.370 $\pm$ 0.004)$\times$ $10^{-11}$ \\
3.55 $\times$ $10^{5}$ & (4.296 $\pm$ 0.003)$\times$ $10^{-11}$ & (3.522 $\pm$ 0.002)$\times$ $10^{-11}$ & (4.015 $\pm$ 0.003)$\times$ $10^{-11}$ \\
4.47 $\times$ $10^{5}$ & (2.323 $\pm$ 0.002)$\times$ $10^{-11}$ & (1.919 $\pm$ 0.002)$\times$ $10^{-11}$ & (2.173 $\pm$ 0.002)$\times$ $10^{-11}$ \\
5.62 $\times$ $10^{5}$ & (1.262 $\pm$ 0.001)$\times$ $10^{-11}$ & (1.045 $\pm$ 0.001)$\times$ $10^{-11}$ & (1.179 $\pm$ 0.001)$\times$ $10^{-11}$ \\
7.08 $\times$ $10^{5}$ & (6.834 $\pm$ 0.008)$\times$ $10^{-12}$ & (5.678 $\pm$ 0.007)$\times$ $10^{-12}$ & (6.373 $\pm$ 0.007)$\times$ $10^{-12}$ \\
8.91 $\times$ $10^{5}$ & (3.695 $\pm$ 0.005)$\times$ $10^{-12}$ & (3.071 $\pm$ 0.005)$\times$ $10^{-12}$ & (3.433 $\pm$ 0.005)$\times$ $10^{-12}$ \\
1.12 $\times$ $10^{6}$ & (2.001 $\pm$ 0.003)$\times$ $10^{-12}$ & (1.671 $\pm$ 0.003)$\times$ $10^{-12}$ & (1.853 $\pm$ 0.003)$\times$ $10^{-12}$ \\
1.41 $\times$ $10^{6}$ & (1.092 $\pm$ 0.002)$\times$ $10^{-12}$ & (9.168 $\pm$ 0.020)$\times$ $10^{-13}$ & (1.014 $\pm$ 0.002)$\times$ $10^{-12}$ \\
1.78 $\times$ $10^{6}$ & (5.947 $\pm$ 0.014)$\times$ $10^{-13}$ & (4.993 $\pm$ 0.013)$\times$ $10^{-13}$ & (5.514 $\pm$ 0.014)$\times$ $10^{-13}$ \\
2.24 $\times$ $10^{6}$ & (3.228 $\pm$ 0.009)$\times$ $10^{-13}$ & (2.705 $\pm$ 0.009)$\times$ $10^{-13}$ & (2.978 $\pm$ 0.009)$\times$ $10^{-13}$ \\
2.82 $\times$ $10^{6}$ & (1.738 $\pm$ 0.006)$\times$ $10^{-13}$ & (1.470 $\pm$ 0.006)$\times$ $10^{-13}$ & (1.610 $\pm$ 0.006)$\times$ $10^{-13}$ \\
3.55 $\times$ $10^{6}$ & (9.301 $\pm$ 0.040)$\times$ $10^{-14}$ & (7.683 $\pm$ 0.036)$\times$ $10^{-14}$ & (8.444 $\pm$ 0.038)$\times$ $10^{-14}$ \\
4.47 $\times$ $10^{6}$ & (4.692 $\pm$ 0.025)$\times$ $10^{-14}$ & (3.902 $\pm$ 0.023)$\times$ $10^{-14}$ & (4.261 $\pm$ 0.024)$\times$ $10^{-14}$ \\
5.62 $\times$ $10^{6}$ & (2.384 $\pm$ 0.016)$\times$ $10^{-14}$ & (1.965 $\pm$ 0.014)$\times$ $10^{-14}$ & (2.133 $\pm$ 0.015)$\times$ $10^{-14}$ \\
7.08 $\times$ $10^{6}$ & (1.159 $\pm$ 0.010)$\times$ $10^{-14}$ & (9.510 $\pm$ 0.090)$\times$ $10^{-15}$ & (1.031 $\pm$ 0.009)$\times$ $10^{-14}$ \\
8.91 $\times$ $10^{6}$ & (5.571 $\pm$ 0.061)$\times$ $10^{-15}$ & (4.529 $\pm$ 0.055)$\times$ $10^{-15}$ & (4.976 $\pm$ 0.058)$\times$ $10^{-15}$ \\
1.12 $\times$ $10^{7}$ & (2.631 $\pm$ 0.037)$\times$ $10^{-15}$ & (2.202 $\pm$ 0.034)$\times$ $10^{-15}$ & (2.321 $\pm$ 0.035)$\times$ $10^{-15}$ \\
1.41 $\times$ $10^{7}$ & (1.265 $\pm$ 0.023)$\times$ $10^{-15}$ & (1.103 $\pm$ 0.022)$\times$ $10^{-15}$ & (1.160 $\pm$ 0.022)$\times$ $10^{-15}$ \\
1.78 $\times$ $10^{7}$ & (6.631 $\pm$ 0.149)$\times$ $10^{-16}$ & (5.818 $\pm$ 0.140)$\times$ $10^{-16}$ & (5.924 $\pm$ 0.141)$\times$ $10^{-16}$ \\
2.24 $\times$ $10^{7}$ & (3.177 $\pm$ 0.092)$\times$ $10^{-16}$ & (2.743 $\pm$ 0.086)$\times$ $10^{-16}$ & (2.787 $\pm$ 0.086)$\times$ $10^{-16}$ \\
2.82 $\times$ $10^{7}$ & (1.651 $\pm$ 0.059)$\times$ $10^{-16}$ & (1.454 $\pm$ 0.056)$\times$ $10^{-16}$ & (1.476 $\pm$ 0.056)$\times$ $10^{-16}$ \\
3.55 $\times$ $10^{7}$ & (8.101 $\pm$ 0.370)$\times$ $10^{-17}$ & (7.048 $\pm$ 0.345)$\times$ $10^{-17}$ & (6.516 $\pm$ 0.331)$\times$ $10^{-17}$ \\
4.47 $\times$ $10^{7}$ & (3.614 $\pm$ 0.220)$\times$ $10^{-17}$ & (3.120 $\pm$ 0.204)$\times$ $10^{-17}$ & (3.038 $\pm$ 0.202)$\times$ $10^{-17}$ \\
5.62 $\times$ $10^{7}$ & (2.146 $\pm$ 0.151)$\times$ $10^{-17}$ & (2.013 $\pm$ 0.146)$\times$ $10^{-17}$ & (1.848 $\pm$ 0.140)$\times$ $10^{-17}$ \\
7.08 $\times$ $10^{7}$ & (9.764 $\pm$ 0.908)$\times$ $10^{-18}$ & (8.343 $\pm$ 0.840)$\times$ $10^{-18}$ & (8.345 $\pm$ 0.840)$\times$ $10^{-18}$ \\
8.91 $\times$ $10^{7}$ & (4.156 $\pm$ 0.528)$\times$ $10^{-18}$ & (4.654 $\pm$ 0.559)$\times$ $10^{-18}$ & (4.086 $\pm$ 0.524)$\times$ $10^{-18}$ \\
1.12 $\times$ $10^{8}$ & (2.694 $\pm$ 0.379)$\times$ $10^{-18}$ & (2.585 $\pm$ 0.371)$\times$ $10^{-18}$ & (2.636 $\pm$ 0.375)$\times$ $10^{-18}$ \\
1.41 $\times$ $10^{8}$ & (1.242 $\pm$ 0.229)$\times$ $10^{-18}$ & (1.286 $\pm$ 0.233)$\times$ $10^{-18}$ & (11.154 $\pm$ 2.174)$\times$ $10^{-19}$ \\
1.78 $\times$ $10^{8}$ & (5.310 $\pm$ 1.337)$\times$ $10^{-19}$ & (5.647 $\pm$ 1.379)$\times$ $10^{-19}$ & (4.922 $\pm$ 1.287)$\times$ $10^{-19}$ \\
\tableline
\tableline
\end{tabular}
\label{tab:03}
\end{center}
\end{table}

%% If the table is more than one page long, the width of the table can vary
%% from page to page when the default \tablewidth is used, as below.  The
%% individual table widths for each page will be written to the log file; a
%% maximum tablewidth for the table can be computed from these values.
%% The \tablewidth argument can then be reset and the file reprocessed, so
%% that the table is of uniform width throughout. Try getting the widths
%% from the log file and changing the \tablewidth parameter to see how
%% adjusting this value affects table formatting.

%% The \dataset{} macro has also been applied to a few of the objects to
%% show how many observations can be tagged in a table.

%% The following command ends your manuscript. LaTeX will ignore any text
%% that appears after it.

\end{document}